\documentclass[12pt]{JHEP}
\usepackage{amsmath,amssymb}

\newcommand{\be}{\begin{equation}}
\newcommand{\ee}{\end{equation}}
\newcommand{\beqa}{\begin{eqnarray}}
\newcommand{\eeqa}{\end{eqnarray}}
\newcommand{\half}{\frac{1}{2}}

\def\med{\frac{1}{2}}
\def\d{\partial}

\def\eeq{\end{equation}}
\def\half{\frac{1}{2}}

\relax

\preprint{CECS-PHY-04/05\\ {\tt hep-th/0403179}}

\title{Localized Intersections of Non-Extremal p-branes and
S-branes}

\author{Jos\'e D. Edelstein$^{\dag \natural}$, Javier Mas$^{\ddag}$
\\
$^{\dag}$Instituto de F\'\i sica de La Plata -- Conicet,
Universidad Nacional de La Plata \\ C.C. 67, (1900) La Plata,
Argentina\\
\medskip
$^{\natural}$Centro de Estudios Cient\'\i ficos (CECS), Casilla
1469, Valdivia,
Chile\\
\medskip
$^{\ddag}$Departamento de F\'\i sica de Part\'\i culas,
Universidad de Santiago de Compostela \\ E-15782 Santiago
de Compostela, Spain\\
\medskip
\email{edels@fisica.unlp.edu.ar}, \quad
\email{jamas@fpaxp1.usc.es}}

\abstract{A class of solutions to Supergravity in 10 or 11
dimensions is presented which extends the non-standard or
semi-local intersections of $Dp$-branes to the case of
non-extremal p-branes. The type of non-extremal solutions involved
in the intersection is free and we provide two examples involving
black-branes and/or  $D\!-\!\bar{D}$ systems. After a  rotation
among the time coordinate and a relatively transverse radial
direction the solutions  admit the interpretation of an
intersection among D-branes and S-branes. We speculate on the
relevance of these configurations  both to study time dependent
phenomena in the AdS/CFT correspondence as well as to construct
cosmological brane-world scenarios within String Theory admitting
accelerating expansion of the Universe.}


\begin{document}

\section{Introduction}

That extremal $p$-brane supergravity solutions correspond to the
macroscopic description of string theory $Dp$-branes is a well
established and fairly understood issue. This neat correspondence
has been very fruitful in understanding aspects like the entropy
of extremal black holes or the AdS/CFT correspondence. These
objects admit a further description as solitons in the tachyonic
field theory of unstable branes \cite{Sen}. Regarding
non-supersymmetric branes, whose stability is not even guaranteed,
the precise identification among the open and closed string theory
point of views is more subtle and disputable. Such is the case of
black-branes, $D\!-\!\bar{D}$ systems or, more recently, a class
of time dependent solutions generically termed $S$-branes
\cite{GS} which have been the subject of intense analysis (see,
for example, \cite{CGG,KMP,BR}). Being time dependent solutions,
it is natural to conjecture that they may represent a macroscopic
description of the rolling tachyon that drives the decay of
unstable branes \cite{GS}. Still this interpretation, mainly due
to the presence of singularities, is a subject of a controversy
\cite{BLW,BW,LePe} which is well beyond the scope of this paper.
Therefore, just to better fix the nomenclature of this article, by
$S$-branes we shall denote the aforementioned class of time
dependent solutions to $D=10,11$ dimensional supergravity. We are
confident that our construction is robust enough to encompass more
general solutions, like the regular ones that are being announced
while we type the last paragraphs of this paper \cite{Adv}.

In the present article we will study configurations admitting the
interpretation of an intersection of $D$-branes and $S$-branes.
Motivations to look for such a system can be gathered from
different arenas. For example, from the rolling tachyon framework,
the picture advocated in \cite{GS} can be generalized to
asymmetric decays along a  spacelike dimension $x$. Say, for
different sign$(x) = \pm 1$ the tachyon field, $T(x,t)$, could
decay in opposite directions of the tachyon potential. The end
process of such a behavior would be a timelike kink, identifiable
with a $Dp$-brane in the far future. Alternatively, one might
think of a stable $Dp$-brane intersecting an unstable
$D(p'+1)$-brane such that a tachyonic timelike kink in the
worldvolume of the latter gives raise to an $Sp'$-brane that
intersects the original stack of $Dp$-branes. In a way or the
other, it is clear that it is meaningful to consider a
configuration that superimposes the tachyonic origin of $D$-branes
and $S$-branes described above. The microscopic meaning of such a
time dependent process deserves investigation, and whatever it is,
one should also expect a macroscopic effective description coming
from supergravity. In a first approximation, at least, the
solutions we will present here have the correct symmetries and
charges to be a good starting point. As well, known solutions
corresponding to different kinds of branes can be recovered as
particular limits of ours.

Leaving the rolling tachyon aside as an inspiring mechanism, one
could just be interested in time dependent solutions to
supergravity. In the literature these have mainly been worked out
in connection with cosmology \cite{Cosmo,Quevetal,MiLo} More
recently, some efforts have been devoted to understand and
overcome the difficulties that prevent to embed an accelerating
cosmology into a higher dimensional consistent framework like $M$
theory \cite{GMN}. Mainly in \cite{WT} the importance of admitting
time dependent compactifications was illustrated in a concrete
example where the compactified manifold had constant negative
curvature. It was a nice observation made in \cite{Ohta,Roy} that
this solution coincides precisely with a neutral supergravity
$SM2$-brane. An $Sp$-brane has $p+1$ Euclidean dimensions. Hence
the $SM2$ yields precisely $3$ dimensions which can account for
the three dimensional space. The remnant outer space needs to be
compactified. While this counting is fine for cosmology, from a
particle physicist point of view there is little support to
considering an $SM2$-brane instead of any other $Sp$-brane or,
say, an $SM5$-brane with $3$ Euclidean longitudinal coordinates
compactified on a torus, or even intersections thereof
\cite{Ohtair,DKI}. While $D$-branes are known to trap gauge
interactions along their worldvolume, it is less clear that
$S$-branes work alike (the corresponding field theories should be
Euclidean). If one is interested in a construction that performs
both tasks, of trapping gauge interactions and having a time
dependence, then an intersection between $D$-branes and $S$-branes
looks promising.

The ideal situation would involve a stack of $D3$-branes
intersecting a space-filling neutral $SD8$-brane. The metric would
exhibit a double warping of the component functions. One in $r$,
the radial distance to the position of the $D3$-branes along the
$SD8$-brane Euclidean nine dimensional  worldvolume. The other in
$t$, the only coordinate outside the $SD8$-brane which is placed
exactly at a fixed time $t= t_0$. In a first approximation we
would like to answer questions like: does a configuration like the
previous one exist? Is the associated cosmology viable? Does it
exhibit late time accelerated expansion? Are there other
interesting examples? What are the general intersection rules and
constraints? How many parameters are contained in the most general
solution? Questions of a different nature also arise such as:  are
there time dependent processes in the quantum field theory living
on the $D3$-branes that are dual to these solutions from the
AdS/CFT point of view?

In order to make some progress along these lines it is convenient
to take advantage from what is known about intersections
\cite{RevInt}, mainly of non-extremal $p$-branes
\cite{CT,AIV,Ohta-ne}. Moreover, we are looking for solutions that
depend on two coordinates $r$ and $t$. These will be the outer
coordinates to the $D$ and $S$-branes respectively. This reminds
of a similar setup affronted by the so called semi-localized (or
non-standard) intersections \cite{SemiLoc1,ETT,SemiLoc2}. However
with the exception of reference \cite{Aref} these intersections
have always dealt with extremal branes. Moreover, the two
coordinates were spacelike radial distances to the intersection
region along the relative transverse directions. Therefore, on top
of generalizing the ansatz to admit non-extremal solutions, we
must include a signature freedom that allows us, at the end, to
trade the time for one of these radial coordinates. In short, the
ansatz we will analyze in the present paper contains the following
line element
\begin{equation} ds^2 =-\eta_T T^2dt^2 + C^2 d\vec z^{\,2}_{ q} +
\eta_X X^2 dx^2 + U^2 d\Omega_{X,\sigma_X}^2 + \eta_Y Y^2 dy^2 +
V^2 d\Omega_{Y,\sigma_Y}^2 + W^2 d\vec w_{p_t}^{\,2} ~.
\label{metric1}
\end{equation}
It has the correct symmetries to represent the intersection of a
pair of branes. Call them $X$ and $Y$. We will refer to them
generically with the index $A= X,Y$. Each p-brane has dimension
$q_A = q+1+p_A$, thus we will also denote them as $q_A$-branes.
$q$ is the dimension of the common Euclidean space where they
intersect, which is spanned by cartesian coordinates
$z_1,...,z_q$.  The coordinates $ x,\theta_X^1,...,\theta_X^{p_X}$
span the space along the $q_X$-brane and transverse relative to
the $q_Y$-brane. Conversely, the $q_Y$-brane has coordinates
$y,\theta_Y^1,...,\theta_Y^{p_Y}$ along the space relatively
transverse to the $X$-brane. The cartesian coordinates
$w_1,...,w_{p_t}$ span the overall transverse space. The angular
variables $\theta_A^1,...,\theta_A^{p_A}$ are coordinates of a
maximally symmetric space with line element
$d\Omega^2_{p_A,\sigma_A}$, {\em i.e.} $R^A_{\theta_A^a\theta_A^b}
= \sigma_A(p_A-1) g_{\theta_A^a\theta_A^b}$, where $\sigma_A=
+1,-1,0$ denotes respectively a spherical, hyperbolic or flat
space. This ansatz must be supplemented with one for the forms
respecting the isometries. All the functions will depend on both
``radial'' relatively transverse coordinates $x$ and $y$. A not
minor assumption will be the factorizability of all the functions
${\cal F}(x,y) = {\cal F}_x(x){\cal F}_y(y)$. (Subindices $x,y$
will generically refer in this paper to functions depending on
those variables.) Eventually we will like to reinterpret one of
them, say $x$, as a time. Therefore we have inserted phases
$\eta_T,\eta_X,\eta_Y = \pm 1$. We will carry them all along the
computations and thus postpone to the end the decision where to
put the timelike signature. Of course, special attention should be
paid to the appearance of imaginary charges and fluxes.

For $\eta_T= \eta_X = \eta_Y=1$, the extremal case analyzed in
\cite{ETT} is recovered in the special limit $T = C, ~X = U $ and
$Y = V$ with $\sigma_X = \sigma_Y = +1$. Then $x$ and $y$ are the
radial distances to the intersection region along  the Euclidean
$p_X+1$ and $p_Y+1$ dimensional relative transverse spaces. In
\cite{Aref} an intersection involving black-branes has been worked
out. It will be reobtained in a later section as a particular case
of our construction. By keeping track of the phases $\eta_A$, the
great generality of our ansatz provides enough room to consider
pairwise partially localized intersections of any of the following
objects: $D$-branes, $E$-branes \cite{Hull}, black-branes,
$S$-branes, $D\!-\!\bar{D}$ systems and Schwarzschild branes
(neutral black-branes).

The paper is organized in the following form. In section 2 we will
obtain the general form of the separated equations of motion. In
obtaining this system the starting ansatz (\ref{metric1}) must be
supplemented by some further constraints. In the following three
sections we shall exhibit particular solutions of the system,
representing intersections of black-branes with black-branes
(section 3), $D\!-\!\bar{D}$ systems (section 4) and $S$-branes
(section 5; in this last case we actually focus in the extremal
limit on the p-brane side). We obtain intersection rules for all
these cases, the full system of algebraic equations that constrain
the parameters of the corresponding solutions and we elaborate on
some aspects of these configurations that might be relevant in the
light of the motivations stated above. Section 6 includes the
discussion and speculation on the relevance of these
configurations to study time dependent phenomena in the AdS/CFT
correspondence as well as to construct cosmological scenarios
within String Theory admitting accelerating expansion of the
Universe. Appendices A and B display, for completeness, details
about the computations that lead to the final expressions given in
the text. Appendix C exhibits a list of explicit solutions to the
algebraic constraints given in section 3.

\section{Ansatz and Separability of the Equations of Motion}

We shall deal with the usual action for a generic dilatonic
Einstein-Maxwell gravity theory in $D$ spacetime dimensions
\begin{equation} S = \frac{1}{16\pi G_D} \int d^D
x \sqrt{-g} \left( R - \frac{1}{2} (\partial \phi)^2 -
\sum_{A=1}^Q \frac{1}{2 n_A !}e^{a_A \phi}F_{n_A}^2 \right) ~,
\label{action}
\end{equation}
The action includes gravity, a dilaton and $Q$ field strengths of
arbitrary form degree $n_A \le D/2$ and coupling to the dilaton
$a_A$. We will eventually set $D = 10, 11$ and choose appropriate
values of $a_A$ to make contact with $IIA$, $IIB$ and $11d$
supergravity theories though there is clearly room to consider
more general cases. The equations of motion can be written in the
following form:
\begin{eqnarray}
&&{R^\mu}_\nu ~=~ \half \d^\mu \phi \d_\nu \phi + \sum_{A=1}^Q
\frac{1}{2n_A!}e^{a_A\phi} \left(n_A F^{\mu\rho_2\ldots\rho_{n_A}}
F_{\nu\rho_2\ldots\rho_{n_A}}- \frac{n_A-1}{D-2} F_{n_A}^2
{\delta^\mu}_\nu \right) ~. ~, \label{einstein} \\
&&\frac{1}{\sqrt{-g}}\d_\mu(\sqrt{-g}\d^\mu)\phi ~=~ \sum_{A=1}^Q
\frac{a_A}{2n_A!}e^{a_A\phi}F_{n_A}^2 ~, \label{dilaton} \\
&& \d_{\mu_1}\left(\sqrt{-g}e^{a_A\phi}F^{\mu_1\ldots\mu_{n_A}}
\right)~=~0 ~, \label{maxwell}
\end{eqnarray}
supplemented with the Bianchi identities
$\d_{[\nu}F_{\mu_1\ldots\mu_{n_A}]}=0$. Also, since we restrict to
pairwise intersections, we shall set $Q=2$. For the metric
functions, the ansatz shown in (\ref{metric1}) will be used.
Concerning the $RR$ forms, we have, in the electric frame, the
following fluxes
\begin{eqnarray}
F^X_{t,1...q\,\theta^1_X \dots \theta^{p_X}_X x\,y} =
\sqrt{\gamma(\theta^a_X)} ~\d_y E^X \label{dflx} \\ F^Y_{t,1...q\,
\theta^1_Y \dots \theta^{p_Y}_Y y\,x} =\sqrt{\gamma(\theta^a_Y)}
~\d_x E^Y\label{dfly}
\end{eqnarray}
where $E^A(x,y)$ are the gauge potentials and
$\gamma(\theta^a_X)$, $\gamma(\theta^a_Y)$ are the volume factors
of the corresponding maximally symmetric spaces. For the case of
magnetic branes, these field strengths must be understood as the
Hodge duals of the relevant ones,
$$
\tilde F_A^{\theta^1_B \dots \theta^{p_B}_B} = \frac{1}{(D -
n_A)!} \frac{e^{-a_A\phi}}{\sqrt{-g}} \epsilon^{\theta^1_B \dots
\theta^{p_B}_B  \mu_1 ...\mu_{n_A} }
F^A_{\mu_1....\mu_{n_A }} ~~ ~~~~ A = X,Y
$$
As mentioned above, the factorization assumption states that
${\cal F}(x,y) = {\cal F}_x(x){\cal F}_y(y)$ for all functions.
Following \cite{ETT} in order to bring the equations of motion
into a manageable form, the following definitions will prove to be
instrumental
\begin{eqnarray}
f_x(x)  &=&x^{-p_X} T_xC^q_x U^{p_X}_x V^{p_Y}_x W^{p_t}_x
X^{-1}_x Y_x \label{funct} \\ f_y(y)  &=&\rule{0mm}{6mm}
y^{-p_Y}T_yC^q_y U^{p_X}_y V^{p_Y}_y W^{p_t}_y X_y Y^{-1}_y
\label{funcs} \\ S^X &=&  \rule{0mm}{8mm}\displaystyle \frac{
W^{2p_t} V^{2p_Y} }{ \sqrt g } e^{\epsilon_X a_X\phi}~ = ~ \frac{
W^{p_t}}{T C^{q} XY} \frac{V^{p_Y}}{U^{p_X} } e^{\epsilon_X
a_X\phi} \label{sx} \\ S^Y &=& \rule{0mm}{7mm} \displaystyle
\frac{W^{2p_t} U^{2p_X}}{\sqrt{-g}} e^{\epsilon_Y a_Y\phi} ~=~ ~
\frac{ W^{p_t} }{T C^{q} XY} \frac{U^{p_X}}{V^{p_Y} }
e^{\epsilon_Y a_Y\phi} \label{sy}
\end{eqnarray}
where $\epsilon_A$ is a positive (negative) sign that accounts for
electric (magnetic) branes. With these ingredients, the diagonal
Einstein  equations read as follows
\begin{eqnarray}
&&\Box \ln T = \Box \ln C = \eta\left( \delta^{Y}_{||}\,
\frac{S^Y}{\sqrt{g}} (\d_x E^Y)^2 +
~\delta^X_{||}\,\frac{S^X}{\sqrt{g}} (\d_y E^X)^2   \right)
\nonumber\\ && \Box \ln X  + \eta_X X^{-2}[ \d_x^2(\ln fx^{p_X}) -
2\d_x\ln X\d_x(\ln fx^{p_X})+ (\d_x\ln T)^2 + q(\d_x\ln C)^2
\nonumber\\ && ~~~~ + p_X(\d_x\ln U)^2 + p_Y(\d_x\ln V)^2 +
p_t(\d_x\ln W_t)^2 + (\d_x\ln Y)^2 - (\d_x\ln X)^2] \nonumber\\
\rule{0mm}{5mm} &&~~~~~~~= ~  \eta\left(
\delta^{Y}_{||}\,\frac{S^Y}{\sqrt{g}} (\d_x E^Y)^2 +
\delta^X_{||}\,\frac{S^X}{\sqrt{g}} (\d_y E^X)^2 \right) - \eta_X
X^{-2}[\med (\d_x\phi)^2  ] \nonumber\\ && \Box \ln Y + \eta_Y
Y^{-2}[ \d_y^2(\ln fy^{p_Y}) - 2\d_y\ln Y\d_y(\ln fy^{p_Y}) +
(\d_y\ln T)^2 + q(\d_y\ln C)^2   \nonumber\\ && ~~~~ + p_X(\d_y\ln
U)^2 + p_Y(\d_y\ln V)^2 + p_t(\d_y\ln W_t)^2 + (\d_y\ln X)^2 -
(\d_y\ln Y)^2  ]\nonumber\\ && ~~~~~~~=~ \eta\left( ~
\delta^{Y}_{||}\,\frac{S^Y}{\sqrt{g}}    (\d_x E^Y)^2 +
\delta^X_{||}\,\frac{S^X}{\sqrt{g}} (\d_y E^X)^2\right) - \eta_Y
Y^{-2}[\med(\d_y\phi)^2]\nonumber\\ &&\Box \ln U  -
\frac{\sigma_X(p_X-1)}{U^2} = \eta\left(  ~
\delta^{Y}_{\perp}\,\frac{S^Y}{\sqrt{g}} (\d_x E^Y)^2 +
\delta^X_{||}\,\frac{S^X}{\sqrt{g}} (\d_y E^X)^2
\right)\nonumber\\ && \Box \ln V  -\frac{\sigma_Y(p_Y-1)}{V^2} =
\eta\left(  ~  \delta^{Y}_{||}\,\frac{S^Y}{\sqrt{g}}    (\d_x
E^Y)^2 + \delta^X_{\perp}\,\frac{S^X}{\sqrt{g}} (\d_y
E^X)^2\right) \nonumber\\ && \Box \ln W =  \eta\left( ~
\delta^{Y}_{\perp}\,\frac{S^Y}{\sqrt{g}} (\d_x E^Y)^2 +
\delta^X_{\perp}\,\frac{S^X}{\sqrt{g}} (\d_y E^X)^2 \right)
\end{eqnarray}
where $\eta \equiv \eta_T\eta_X\eta_Y$, and
$\delta^A_{||},~\delta^A_{\perp}$ are the typical exponents
entering the harmonic superposition rule
\begin{equation}
\delta^A_{||} =\left(\frac{D-q_A-3}{2(D-2)}\right)~~~~~;~~~~~
\delta^A_{\perp} =  \left(\frac{-q_A-1}{2(D-2)}\right)~.
\end{equation}
The D'Alembertian is given by
\begin{equation}
\Box = \frac{1}{\sqrt{|G|}}\d_M(\sqrt{|G|}\d^M) =
\frac{\eta_X}{X^{2}}[\d_x\ln(fx^{p_X})\d_x + \d_x^2] +
\frac{\eta_Y}{Y^{2}}[\d_y\ln (fy^{p_Y}) \d_y + \d_y^2].
\end{equation}
In addition, there is a single off diagonal equation
\begin{eqnarray}
R_{xy}& = & -\d_{xy}(\ln fx^{p_X}y^{p_Y}) + \d_x\ln Y(\d_y\ln
fy^{p_Y}) + \d_y\ln X(\d_x\ln fx^{p_X}) \nonumber\\ && -\d_x\ln
T\d_y \ln T -q\d_x\ln C\d_y\ln C - p_X\d_x\ln U\d_y\ln U
-p_Y\d_x\ln V\d_y\ln V \nonumber\\ && -p_t\d_x\ln W\d_y\ln W
+\d_x\ln Y\d_y\ln Y + \d_x\ln X\d_y\ln X - 2\d_x\ln Y\d_y\ln X
\nonumber\\ & = & ~\med \d_x\phi \d_y\phi ~, \label{offe}
\end{eqnarray}
where the first term straightforwardly vanishes from the
factorization ansatz. As for the dilaton equation, we obtain
\begin{equation}
\Box\phi =- \frac{\eta}{2} \frac{\epsilon_X a_X S^X}{\sqrt{-g}}
(\d_yE^{X})^2 - \frac{\eta}{2} \frac{\epsilon_Y a_Y
S^Y}{\sqrt{-g}} (\d_x E^Y)^2
\end{equation}
Finally, the equations of motion satisfied by the gauge potentials
are compactly written as
\begin{equation}
\begin{array}{rcl}
\d_x \left(S^X(\d_y E^X)\right) &=& 0 \\
\d_y \left(S^X(\d_y E^X)\right) &=& 0
\end{array} ~~~ ; ~~~ \begin{array}{rcl}
\d_y \left(S^Y(\d_x E^Y)\right) &=& 0 \\
\d_x \left(S^Y(\d_x E^Y)\right) &=& 0
\end{array} \label{maxw}
\end{equation}
We shall start from here solving the full system. Invoking  factorizability
these equations can be integrated to give
\begin{equation}
\begin{array}{rcl}
S^X_x E^X_x &=& l_X \\
S^X_y(\d_y E^X_y) &=& c_X
\end{array} ~~~ ; ~~~ \begin{array}{rcl}
S^Y_y E^Y_y &=& l_Y \\
S^Y_x(\d_x E^Y_x) &=& c_Y
\end{array} \label{maxs}
\end{equation}
where $l_A$ and $c_A$ are constants. In passing by, and for later
use, let us mention that the product $l_A c_A$ gives basically the
RR-charge (density per unit W space volume)
\begin{eqnarray}
{\cal Q}_X &=& \int d\theta^1_Y...d\theta^{p_Y}_Y \tilde
F^X_{\theta^1_Y..\theta^{p_Y}_Yw^1...w_{p_t}}  = \eta \, l_Xc_X
\,\Omega_{p_Y}  \label{cax}\\ {\cal Q}_Y &=& \int
d\theta^1_X...d\theta^{p_X}_X \tilde
F^Y_{\theta^1_X..\theta^{p_X}_X w^1...w_{p_t}}  = \eta  \,
l_Yc_Y\, \Omega_{p_X}\label{cay}
\end{eqnarray}
with $\Omega_{p_A}
=2\pi^{\frac{p_A+1}{2}}/\Gamma(\frac{p_A+1}{2})$. In order to
proceed, let us insert the last two expressions in (\ref{maxs})
into the Einstein Equations. After multiplying by $X_y^2 Y_x^2$ it
is easy to recognize the conditions that must be imposed in order
to achieve separability of the system. They are:
\begin{eqnarray}
\displaystyle \frac{ X_y^2 }{S^Y_y \sqrt{g}_y} &=&  X_y^2
W_y^{-2p_t} U_y^{-2p_X} e^{- \epsilon_Y a_Y\phi_y} ~=~ 1
\label{spu}
\\ \displaystyle \frac{Y_x^2}{S^X_x\sqrt{g}_x}    &=&  Y_x^2
W_x^{-2p_t} V_x^{-2p_Y} e^{- \epsilon_X a_X\phi_x}~=~ 1 ~.
\label{spd}
\end{eqnarray}
Notice that this set of conditions is the same as in \cite{ETT}
for the critical case and it is scalar under reparametrizations of
either $x$ or $y$. Still, at this stage it is not clear whether
they overdetermine the system or not, so they will have to be
checked for consistency a posteriori. Now, it is not hard to
realize that (\ref{spu}) and (\ref{spd}) are not sufficient in
order to implement the separation of variables. From the angular
$R^{\theta_X}{_{\theta_X}}$ and $R^{\theta_Y}{_{\theta_Y}}$
equations for $\ln U$ and $\ln V$, it is immediate to see that the
following conditions are also needed:
\begin{equation}
Y_x = V_x ~~~~~~~ ; ~~~~~~~ X_y = U_y \label{dkmu}
\end{equation}
This constraint (which is not necessary for $p_X$ or $p_Y=1$) is a
natural one stating that the metric functions along  each brane
behave alike when separating along the relative transverse radial
direction. After all these considerations, the separated equations
adopt the following succinct form,
\begin{eqnarray}
&& (x^{p_X}f_x  (\ln T_x)')'= d^Y_{||}{E^Y_x}'
    + \kappa_T f_x x^{p_X}  \nonumber\\
&& (x^{p_X}f_x  (\ln C_x)')'=   d^Y_{||}{E^Y_x}'
    + \kappa_C f_x x^{p_X}\nonumber\\
&& (\ln X_x)'' - (\ln f_x x^{p_X})' (\ln X_x)'-{(\ln X_x)'}^2  +
    (\ln f_x x^{p_X})''+ (\ln T_x)'{^2} + q{(\ln C_x)'}^2 \nonumber\\
&& ~~~~~~~~+p_X{(\ln U_x)'}^2 + (p_Y+1){(\ln V_x)'}^2 + p_t {(\ln
W_x)'}^2 = ~-\frac{1}{2}{\phi_x'}^2 + d^Y_{||} \frac{{E^Y_x}'}{
f_x \, x^{p_X}} + \kappa_X \nonumber\\
&& (x^{p_X}f_x  (\ln Y_x)')'= d^Y_{||} {E^Y_x}'
    + \kappa_Y f_x x^{p_X}\label{eqse}\\
&& (x^{p_X}f_x  (\ln U_x)')'= d^Y_{\perp} {E^Y_x}' +
    \eta_X\sigma_X(p_X-1)\left({X_x}/{U_x}\right)^2 x^{p_X} f_x
    + \kappa_U f_x x^{p_X}\nonumber\\
&& (x^{p_X}f_x   (\ln V_x)')'=   d^Y_{||}{E^Y_x}'
    + \kappa_V f_x x^{p_X} \nonumber\\
&& (x^{p_X}f_x   (\ln W_x)')'=   d^Y_{\perp}{E^Y_x}'
    + \kappa_W f_x x^{p_X}\nonumber\\
&& (x^{p_X}f_x   \phi_x')' = -e^Y {E^Y_x}'
    + \kappa_\phi f_x x^{p_X}   \nonumber
\end{eqnarray}
and, conversely,
\begin{eqnarray}
&& (y^{p_Y} f_y (\ln T_y)')'=   d^X_{||}{E^X_y}'
    - \kappa_T f_y y^{p_Y}  \nonumber\\
&& (y^{p_Y} f_y (\ln C_y)')'=   d^X_{||}{E^X_y}'
    - \kappa_C f_y y^{p_Y}  \nonumber\\
&& (\ln Y_y)'' - (\ln f_y y^{p_Y})' (\ln Y_y)'-{(\ln Y_y)'}^2  +
(\ln f_y y^{p_Y})''+ (\ln T_y)'{^2} + q{(\ln C_y)'}^2 \nonumber\\
&& ~~~~~~~~+ p_Y{(\ln V_y)'}^2 +  (p_X+1){(\ln U_y)'}^2 + p_t
{(\ln W_y)'}^2 = ~-\frac{1}{2}{\phi_y'}^2 + d^X_{||}
\frac{{E^X_y}'}{ f_y \,y^{p_Y}} - \kappa_Y \nonumber\\
&& (y^{p_Y} f_y (\ln X_y)')'=   d^X_{||} {E^X_y}'
    - \kappa_X f_y y^{p_Y} \label{eqsd}\\
&& (y^{p_Y} f_y (\ln V_y)')'= d^X_{\perp} {E^X_y}' +
\eta_Y\sigma_Y(p_Y-1)\left({Y_y}/{V_y}\right)^2 y^{p_Y} f_y
    - \kappa_U f_y y^{p_Y}\nonumber\\
&& (y^{p_Y} f_y  (\ln U_y)')'= d^X_{||}{E^X_y}'
    - \kappa_V f_y y^{p_Y} \nonumber\\
&& (y^{p_Y} f_y   (\ln W_y)')'=   d^X_{\perp}{E^X_y}'
    - \kappa_W f_y y^{p_Y}\nonumber\\
&& (y^{p_Y} f_y   \phi_y')' = -e^X {E^X_y}'
    - \kappa_\phi f_y y^{p_Y}
\nonumber
\end{eqnarray}
where we have defined, for convenience, the following constants
\begin{equation}
d^A_{||} = \tilde c_A\delta^{A}_{||} ~~~;~~~  d^A_{\perp} = \tilde
c_A \delta^{A}_{\perp}~~~;~~~ e^A = \med \tilde c_A
\epsilon_A a_A~~~~;~~~~
\tilde c_A = \eta_T\eta_Al_A^2 c_A~~~~ A= X,Y \label{thj}
\end{equation}
and the tilde denotes derivatives with respect to the only
argument of each function. The original system has been
successfully separated. There remains however a single equation
that mixes both solutions:  the off diagonal Einstein
equation (\ref{offe}), or,
\begin{eqnarray}
&& (\ln Y_x)'(\ln f_y y^{p_Y})' + (\ln X_y)'(\ln f_x x^{p_X})' -
(\ln T_x)'(\ln T_y)' - q(\ln C_x)'(\ln C_y)' \nonumber\\ &&~~ -
p_X(\ln U_x)'(\ln U_y)' - p_Y(\ln V_x)'(\ln V_y)' - p_t(\ln
W_x)'(\ln W_y)'\nonumber\\ &&~~~~~ + (\ln Y_x)'(\ln Y_y)' + (\ln
X_x)'(\ln X_y)' -2(\ln Y_x)'(\ln X_y)' ~=~ \med \phi_x'\phi_y'
\label{offde}
\end{eqnarray}
Since this equation does not contain second derivatives, it
typically will impose constraints among the parameters and
integration constants. The above system of equations contains
contributions that are proportional to  the separation constants
$\kappa_\alpha$. These are genuine to the present setup, revealing
the fact that we are dealing with an intersection.  Although these
constants may surely lead to interesting physics, they make the
analytic integration far more involved. Therefore, at this stage,
we shall set them all equal to zero, $\kappa_\alpha = 0$, and
defer a more general analysis for a later study. We will see that,
even in this simplifying case, there is still a rich structure for
the solutions of this system. With $\kappa_\alpha = 0$, the
separated systems (\ref{eqse}) or (\ref{eqsd}) are nothing but a
slight generalization of the one for a single $q_A$-brane, where
the worldvolume global symmetry has been broken by the presence of
the other brane (generically this just brings in additional
integration constants subjected to some constraints).

The next step would be simply to specify the reader's favorite
reduced ansatz, as well as a choice of $x$ and $y$ coordinates.
Such choice needs not even be symmetric for both branes, as we
shall show in a particular example. The presence of the other
brane is still reflected in the off diagonal Einstein equation.
Ideally one would like to borrow the most general p-brane solution
from the literature, and see what constraints arise when coupling
them through equation (\ref{offde}). In this paper we shall
examine two cases. First we will select a generalized black-brane
ansatz for both intervening branes. This case includes the
solutions found in \cite{Aref}. We provide general formulas for
the integrated expressions and perform a full scan of the possible
intersections.  We will find that the presence of additional
integration constants is essential to admit intersections where
both horizons are nonzero. As a second example, we will affront an
asymmetric ansatz where, while we keep for the $X$-brane the same
black-brane-like solution, we use for the $Y$-brane the three
parameter solution that has been obtained recently in connection
with $S$-brane supergravity solutions \cite{CGG,Ohtair}. It is
related by a coordinate transformation to the one obtained in
Ref.\cite{ZZ} (see also \cite{IvMe}), where it was also found as
part of a more general four parameter family (see also the
appendix of \cite{Sk} for another approach to a general solution).
In an appealing analysis, this three parameter solution was
identified with the supergravity description of a non BPS system
of $N$ $Dp$-branes coinciding with $\bar N$ $\bar{Dp}$-branes
\cite{BMO}. The parameters involved characterize the mass, the net
charge (for $N\neq \bar N$) and the vacuum expectation value of
the tachyon. Therefore we will refer to this solution as the
$D\!-\!\bar{D}$ system (see  also the recent papers \cite{GLC,LuR}
and references therein).

Importantly enough, we will keep track of the phases $\eta_A$ in
order to see what possibilities are at hand to select, say, $x$ as
a time. From this, we will be able to present a family of
solutions that represent intersections of $D$-branes and
$S$-branes. Furthermore, the great generality of our ansatz allows
to consider pairwise partially localized intersections of any of
the following objects: $D$-branes, $E$-branes, black-branes,
S-branes, $D\!-\!\bar{D}$ systems.

\section{Intersecting Black-Branes }

In this section we are going to present a family of solutions
describing partially localized intersections of black-branes. We
collect some of the details of the integration of this system in
Appendix A. In considering intersections among black-branes, one
is naturally led to the following ansatz
\begin{eqnarray}
&&T_x=C_x f_x^{1/2}~~~;~~~X_x = \tilde U_x f_x^{-1/2}~~~
;~~~ U_x= x \tilde U_x~~~;~~~ Y_x = V_x\label{anu}\\
&&T_y=C_y f_y^{1/2}~~~;~~~Y_y ~=  \tilde V_y f_y^{-1/2}~~~ ;~~~
V_y \,= y\tilde V_y ~~~;~~~X_y =   U_y \label{ando}
\end{eqnarray}
which leads to the Schwarzschild-like line element (after omitting
tildes)
\begin{eqnarray}
ds^2 &=& C^2 \left(- \eta_Tf_x f_y dt^2 +   d\vec z^{\,2}_q
\right) + U^2 \left(\eta_X f_x^{-1} dx^2 + x^2
d\Omega_{p_X,\sigma_X}^2 \right) + V^2 \left( \eta_Y f^{-1}_y dy^2
+ y^2 d\Omega_{p_Y,\sigma_Y}^2 \right)  \nonumber\\
&&+ W^2 d\vec w_{p_t}^{\,2}
\end{eqnarray}
Inserting equations (\ref{anu})--(\ref{ando}) in
(\ref{funct})--(\ref{funcs}) result in the following gauge
choices:
\begin{eqnarray}
1 = C^{q+1}_x U^{p_X-1}_x V^{p_Y+1}_x W^{p_t}_x   \label{gauu}\\
1 = C^{q+1}_y U^{p_X+1}_y V^{p_Y-1}_y W^{p_t}_y  \label{gaud}
\end{eqnarray}
It turns out that a solution exists if and only if $\eta_A\sigma_A
= +1$, {\em i.e.} $d\Omega^2_{p_A,\sigma_A}$ being a spherical
(hyperbolic) line element if and only if $x_A$ is a spacelike
(timelike) coordinate (see Appendix A for details). The solution
can be obtained in terms of two harmonic functions on each side
\begin{eqnarray}
f_x = 1-\frac{2\mu_Y}{x^{p_X-1}} ~~~~~~&;&~~~~~
f_y = 1-\frac{2\mu_X}{y^{p_Y-1}} \label{efes} \\
H_x = 1 + \frac{Q_Y}{x^{p_X-1}}~~~~~~&;&~~~~~~
H_y = 1 + \frac{Q_X}{y^{p_Y-1}} \label{aches}
\end{eqnarray}
The quantities $Q_A$ are, as usual, proportional to the number of
$q_A$-branes whereas $\mu_A$ are the non-extremal black-brane
parameters. The final expression for the metric functions and the
dilaton are given by
\begin{equation}
\begin{array}{rcl}
C_x &=& f_x^{\frac{c^Y_C}{2\mu_Y(p_X-1)}} ~
    H_x^{-\frac{D-q_Y-3}{\Delta_Y}} \\
U_x &=& f_x^{\frac{c^Y_U+2\mu_Y}{2\mu_Y(p_X-1)}} ~
    H_x^{\frac{q_Y+1}{\Delta_Y}} \\
V_x &=& f_x^{\frac{c^Y_V}{2\mu_Y(p_X-1)}} ~
    H_x^{-\frac{D-q_Y-3}{\Delta_Y}} \\
W_x &=& f_x^{\frac{c^Y_W}{2\mu_Y(p_X-1)}} ~
    H_x^{\frac{q_Y+1}{\Delta_Y}} \\
e^{\phi_x} &=& f_x^{\frac{c^Y_\phi}{2\mu_Y(p_X-1)}} ~
    H_x^{\frac{\epsilon_Ya_Y(D-2)}{\Delta_Y}}
\end{array}
~~~~~~;~~~~~~
\begin{array}{rcl}
C_y &=&  f_y^{\frac{c^X_C}{2\mu_X(p_Y-1)}} ~
    H_y^{-\frac{D-q_X-3}{\Delta_X}} \\
U_y &=& f_y^{\frac{c^X_U}{2\mu_X(p_Y-1)}} ~
    H_y^{-\frac{D-q_X-3}{\Delta_X}}  \\
V_y &=& f_y^{\frac{c^X_V+2\mu_X}{2\mu_X(p_Y-1)}} ~
    H_y^{\frac{q_X+1}{\Delta_X}}  \\
W_y &=& f_y^{\frac{c^X_W}{2\mu_X(p_Y-1)}} ~
    H_y^{\frac{q_X+1}{\Delta_X}}  \\
e^{\phi_y} &=& f_y^{\frac{c^X_\phi}{2\mu_X(p_Y-1)}} ~
    H_y^{\frac{\epsilon_X a_X(D-2)}{\Delta_X}}
\end{array}
\label{hwq}
\end{equation}
whereas for the gauge potentials (\ref{dflx})--(\ref{dfly}) we
obtain
\begin{eqnarray}
E^X  &=&E^X_x E^X_y = \eta_T\eta_X
\sqrt{\eta_T\eta_X\,\frac{Q_X}{(Q_X+2\mu_X)}\frac{2(D-2)}{\Delta_X}
} \,x^{p_X} f_x^{\chi_Y} H_x^{\frac{2(D-2)}{\Delta_Y}}
~\frac{f_y}{H_y} \label{fly}\\
E^Y &=& E^X_x E^X_y =  \eta_T\eta_Y
\sqrt{\eta_T\eta_Y\,\frac{Q_Y}{(Q_Y+2\mu_Y)}\frac{2(D-2)}{\Delta_Y}
}~ ~\frac{f_x}{H_x}~\,y^{p_Y} f_y^{\chi_X}
H_y^{\frac{2(D-2)}{\Delta_X}} \label{flx}
\end{eqnarray}
where the values of the constants $\Delta_A$ and    $\chi_A,~(A = X,Y)$
are given by
\begin{eqnarray}
\Delta_A &=& (q_A+1)(D-q_A-3)+ \med  (D-2) a_A^2  \\
\chi_A &=&  \frac{-1}{\mu_A(p_B-1)}\left(p_t c^A_W + p_A c^A_V -
(c^A_U+2\mu_A) + \med \epsilon_B a_B c^A_\phi\right) ~~~~A\neq B
\end{eqnarray}
For completeness, we have explicitly exhibited the phases $\eta_A$
in the expressions for the forms. We see that if we wanted to
exchange the signatures of, say, $t$ and $x$ by setting $\eta_T =
\eta_X = -1$ in an extremal solution the gauge potential $E^Y$
would become imaginary for $\eta_Y = +1$. This is the case of the
so-called $E$-branes first discussed by Hull \cite{Hull}. However,
for a non zero value of the non-extremality parameter $\mu_Y$,
still a real flux is possible by doing either $Q_Y$ or $\mu_Y$
negative in a certain interval, thus at the price of making the
singularity naked. This seems to agree with the observation in
\cite{BR} that just non-extremal $S$-branes (in this case the
Y-brane) have a chance to admit real charges. The off diagonal
Einstein equation constrains, as usual, the allowed intersections,
through the same equation that was found for extremal solutions
\cite{ETT} \footnote{We would like to remark that this
intersection rule would not have emerged if we had considered,
from the very beginning, that one of the intervening branes was
neutral (Schwarzschild branes). In such a case, there is no
intersection rule and solutions can be found roughly for any
values of $q_X$ and $q_Y$.}
\begin{equation}
q+3 = \frac{(q_X+1)(q_Y+1)}{(D-2)} -
\frac{1}{2}\epsilon_X\epsilon_Y a_X a_Y ~.\label{intr}
\end{equation}
For completeness, we summarize here the general solution to this
equation codified as $(q| X, Y|p_t)_D$,
\begin{eqnarray}
(1|M5,M5|1)_{11} && \label{m5m5}\\
(1|NS5,NS5|0)_{10} &&\label{nsns}\\
(q_X-3|Dq_X, NS5|1)_{10}&~~~& 3\leq q_X \leq 7 \label{dns}\\
(\frac{q_X+q_Y}{2}-4|Dq_X,Dq_Y|5-\frac{q_X+q_Y}{2})_{10}
&~~~~&8\leq q_X+q_Y\leq 10  \label{dcd}
\end{eqnarray}
where, in these expressions, we should remind that we are
generally talking about non-extremal intersecting branes. In
(\ref{hwq}) there are twelve parameters $\mu_A, c^A_C, c^A_U,
c^A_V,c^A_W, c^A_\phi, ~ (A= X,Y)$, subjected to nine constraints.
Two of them arise from the gauge fixing condition
(\ref{gauu})--(\ref{gaud})
\begin{eqnarray}
&& (q+1)c^Y_C  +(p_X-1)(2\mu_Y+c^Y_U) + (p_Y+1) c^Y_V +
    p_t c^Y_W = 0 \label{kki}\\
&& (q+1)c^X_C  +(p_Y-1)(2\mu_X+c^X_V) +  (p_X+1) c^X_U +
    p_t c^X_W = 0 ~. \label{nks}
\end{eqnarray}
In the course of the integration the next four equations appear
(see Appendix A)
\begin{eqnarray}
&(p_X-1) {c^Y_U}^2+ 2\mu_Y (p_X-1)(c^Y_U+c^Y_C) +
(q+1) {c^Y_C}^2 + (p_Y+1){c^Y_V}^2 + p_t {c^Y_W}^2 +
\med {c^Y_\phi}^2 =0 & \label{ghju} \\
&(p_Y-1) {c^X_V}^2+ 2\mu_X (p_Y-1)(c^X_V+c^X_C) + (q+1) {c^X_C}^2
+ (p_X+1){c^X_U}^2 + p_t {c^X_W}^2 + \med {c^X_\phi}^2 =0 &
\label{ghjd}
\\
& 2(p_X-1)(2\mu_Y+c^Y_U)+2p_tc^Y_W+\epsilon_Y a_Yc^Y_\phi = 0 &
\label{gud}
\\ & 2(p_Y-1)(2\mu_X+c^X_V)+2p_tc^X_W+\epsilon_X a_Xc^X_\phi = 0
&\label{gun}
\end{eqnarray}
From these, the first two are unavoidable. The second pair,
however, is just a restriction on the phase space that allows to
express the solution to the gauge potentials $E^A$ in terms of
harmonic functions (\ref{efes})--(\ref{aches}). Finally, from the
off diagonal Einstein equation (\ref{offde}) we obtain the
following three constraints \footnote{For the case in which $X$
($Y$) is a neutral brane, the constraint (\ref{xyd})
(alternatively (\ref{xyu})) would be absent.}
\begin{eqnarray}
&p_t c^Y_W + (p_Y-1) c^Y_V + \med \epsilon_X a_X c^Y_\phi = 0 &
\label{xyu}
\\ &p_t c^X_W + (p_X-1) c^X_U + \med \epsilon_Y a_Y c^X_\phi = 0 &
\label{xyd}
\\
\rule{0mm}{5mm}
&\mu_X\mu_Y(p_X-1)(p_Y-1)+\mu_X(p_Y-1)(c^Y_V+c^Y_C) +
\mu_Y(p_X-1)(c^X_U+c^X_C)~~~~~~~~~~~~& \nonumber \\ &~~~~~~~+
(q+1)c^X_Cc^Y_C + (p_X-1)c^X_Uc^Y_U + (p_Y-1)c^Y_Vc^X_V +p_t c^Y_W
c^X_W + 2c^X_Uc^Y_V + \med c^Y_\phi c^X_\phi =0 ~~~~~~~&
\label{itsc} \rule{0mm}{4mm}
\end{eqnarray}
Let us pause here to comment about the generic features that
solutions to this algebraic system exhibit. First of all, notice
that there is always a simple solution of the first eight
constraints given by $c^A_\alpha = 0$ for all $\alpha$ except
$c^Y_U = -2\mu_Y$ and $c^X_V = -2\mu_X$. With this, the last
equation (\ref{itsc}) reads just $(p_X-1)(p_Y-1)\mu_X\mu_Y = 0$.
Hence one of the two branes has to be set to extremality, either
$\mu_X=0$ or $\mu_Y = 0$. Within this class of solutions, the
non-extremalization follows the simple recipe that was found  for
the case of standard intersections in \cite{CT}. In summary,
generically we always find the two solutions in $D=10$
\begin{eqnarray}
(\mu_X=  0)~~~~~ds^2 &=& H_x^{\frac{q_X+1}{8}}
H_y^{\frac{q_Y+1}{8}} [H_x^{-1} H_y^{-1} (-f_x dt^2 + d\vec
z_q^{\,2}) + H_y^{-1} (f_x^{-1} dx^2 + x^2 d\Omega_{p_X,1}^2)
\nonumber\\ && ~+~ H_x^{-1} (dy^2 + y^2 d\Omega_{p_Y,1}^2) + d\vec
w^{\,2}_{p_t}]
\nonumber\\
(\mu_Y=  0)~~~~~ds^2 &=& H_x^{\frac{q_Y+1}{8}}
H_y^{\frac{q_X+1}{8}} [H_x^{-1} H_y^{-1}(-f_y dt^2 + d\vec
z_q^{\,2}) + H_y^{-1}(  dx^2 + x^2 d\Omega_{p_X,1}^2) \nonumber\\
&& ~+~ H_x^{-1} (f_y^{-1} dy^2 +y^2 d\Omega_{p_Y,1}^2) + d\vec
w^{\,2}_{p_t}]
\end{eqnarray}
and the corresponding ones in $D=11$ for (\ref{m5m5}) with
$(q_A+1)/8$ replaced by $2/3$. We will refer to these as the
``minimal non-extremal'' deformations. However the system
(\ref{kki})--(\ref{itsc}) admits also sometimes non-minimal
extensions. In all the cases listed in (\ref{dns}), except for
$q_X=3$, the solution depends on three parameters, two of which
can be taken as $\mu_X,\mu_Y$. In all the remaining cases in
$D=10$, notice that either   $q=0$ or $p_t = 0 $ and there is a
reduction in the number of parameters (since correspondingly
$c^Y_C$, $c^X_C$ or $c^Y_W$, $c^X_W$ are absent). Also the case
$(1|M5,M5|1)$ has no dilaton and there are no $c^Y_\phi,c^X_\phi$.
These situations never occur simultaneously (except for the cases
$(0|D3,D5|1)$ and $(0|D3,NS5|1)$ where, nevertheless, there is
still a one parameter solution at the end), so in all these cases
there are nine equations for ten variables, and we would expect a
one parameter family of solutions. Solving the first eight
constraints (\ref{kki})-(\ref{xyd}) for all the constants except
$\mu_X$ and $\mu_Y$, yields typically four solutions which one can
introduce into the quadratic equation (\ref{itsc}), which then
collapses to an identity of the sort $A ~\mu_X\mu_Y = 0$ where $A$
is a numerical constant specific to each case.  If $A\neq 0$ this
equation implies, as before, either $\mu_X=0$ or $\mu_Y=0$.
However, frequently, two out of the four solutions have $A = 0$
and, therefore, automatically solve the last constraint. In these
cases both $\mu_X$ and $\mu_Y$ are free and we obtain a
$2$-parameter family of solutions.

Summarizing, the constraints are always solved if one of the
constituents is $BPS$. Most cases also admit an intersection among
two non-extremal branes. Also notice that for each $D3$ or
$M5$-brane involved in the intersection, we must skip the
corresponding $c^A_\phi$ as well. We have summarized, in Appendix
$C$, the complete expressions for all possible intersections. Two
of them agree with the ones found in \cite{Aref}. One can see
that, whereas the $H_x$ and $H_y$ follow a harmonic superposition
rule, there is no such a pattern to encode the way the $f$
functions appear, aside from the interesting fact that always one
of the branes is either $BPS$ or its ``minimal non-extremal''
deformation. It is possible, at this stage, to choose different
values for the phases $\eta_A$ to obtain intersections involving
$S$-branes. However, the ansatz we are dealing with does break
$SO(1,q)$ isometries in the intersection. This would lead to
non-isotropic $S$-branes. We are not going to study those
configurations in this paper. Let us only mention that the kind of
solutions involving non-isometric $S$-branes includes a family of
completely regular solutions that might deserve further study
\cite{KMP}.

\section{Intersecting Black-branes and $D\!-\!\bar{D}$ system}

As mentioned above, by $D\!-\!\bar{D}$ system we refer to a
non-extremal solution that does not break the $SO(1,q)$ symmetry
of the worldvolume. A general solution of this kind can be found
in \cite{ZZ,BMO}. This solution can be nicely fitted within our
approach. However, we prefer to borrow the solution given in
\cite{CGG,Ohta} which is physically equivalent since both are
related by a change of coordinates. The reason for this preference
is twofold. On one side it is much simpler, and on the other it
will allow us, in the upcoming section, to relate this
configuration to another one which can be interpreted as an
intersection of $D$-branes with $S$-branes.

To fix up the notation we will chose $X$ to be the black-brane and
$Y$ the $D\!-\!\bar{D}$ system. For $X$ we adopt the  same ansatz
as in the previous section, therefore its integration follows
closely the steps given in Appendix A after replacing  $x \to y$,
$X \to Y$ and $U \to V$.  For the $Y$-brane, instead, we impose
$T_x = C_x$ which implements the desired $SO(1,q)$ symmetry (as
far as the $Y$-brane is concerned). The line element will acquire
the following mixed structure
\begin{eqnarray}
ds^2 &=& C_x^2 C_y^2\left(-\eta_T f_y dt^2 + d\vec z_q^{\,2}
\right) + U_y^2\left( \eta_X X_x^2 dx^2 + U_x^2
d\Omega^2_{p_X,\sigma_X} \right) + V^2_xV_y^2( \eta_Y f_y^{-1}dy^2
+ y^2 d\Omega_{p_Y,\sigma_Y}^2) \nonumber\\ && ~~~~~ +~ W^2 d \vec
w^2_{\,p_t} \label{seca}
\end{eqnarray}
The $SO(1,q)$ symmetry is broken by the black-brane $X$ through
the non-extremal function $f_y$. Following \cite{CGG,Ohta} the $x$
coordinate gauge fixing that leads to the solutions we are
interested in, is given by $ f_x \,x^{p_X} = 1$, or, looking at
(\ref{funct}), by $1 = C^{1+q}_x U^{p_X}_x V^{p_Y+1}_x W^{p_t}_x
X^{-1}_x.$ With this, one notices that the left hand side of the
Einstein equations (\ref{eqse}) simplifies considerably. Following
these papers it is fairly straightforward to carry out the
integration which the reader can find for completeness in Appendix
B (with the careful inclusion of the phases $\eta_A$). The
solution reads:
\begin{equation}
\begin{array}{rcl}
C_x &=& e^{c^Y_C x } ~
    K_x^{-\frac{D-q_Y-3}{\Delta_Y}} \\
\rule{0mm}{7mm} X_x &=& e^{ c^Y_X x+ p_Xg(x)} ~
    K_x^{\frac{q_Y+1}{\Delta_Y}} \\
\rule{0mm}{7mm} U_x &=& e^{ c^Y_U x+g(x)} ~
    K_x^{\frac{q_Y+1}{\Delta_Y}} \\
\rule{0mm}{7mm}  V_x &=& e^{c^Y_V x } ~
    K_x^{-\frac{D-q_Y-3}{\Delta_Y}} \\
\rule{0mm}{8mm} W_x &=& e^{c^Y_W x } ~
    K_x^{\frac{q_Y+1}{\Delta_Y}} \\
\rule{0mm}{8mm} e^{\phi_x} &=& e^{c^Y_\phi x} ~
    K_x^{\frac{(D-2)a_Y\epsilon_Y}{\Delta_Y}}
\end{array}
~~~~~~~~~~~~~
\begin{array}{rcl}
C_y &=&  f_y^{\frac{c^X_C}{2\mu_X(p_Y-1)}} ~
    H_y^{-\frac{D-q_X-3}{\Delta_X}}  \\
\rule{0mm}{6mm}
X_y &=& U_y  \\
U_y &=& f_y^{\frac{c^X_U}{2\mu_X(p_Y-1)}} ~
    H_y^{-\frac{D-q_X-3}{\Delta_X}}  \\
V_y &=& f_y^{\frac{c^X_V+2\mu_X}{2\mu_X(p_Y-1)}} ~
    H_y^{\frac{q_X+1}{\Delta_X}}  \\
W_y &=& f_y^{\frac{c^X_W}{2\mu_X(p_Y-1)}} ~
    H_y^{\frac{q_X+1}{\Delta_X}}  \\
e^{\phi_y} &=& f_y^{\frac{c^X_\phi}{2\mu_X(p_Y-1)}} ~
    H_y^{\frac{\epsilon_X a_X(D-2)}{\Delta_X}}
\end{array}
\label{sys}
\end{equation}
with $f_y, H_y$ as given in (\ref{efes}) and (\ref{aches}),
$\eta_Y \sigma_Y = +1$, and
\begin{equation}
K_x =  \frac{1}{2}(e^{2\gamma_Y(x-x_Y)}-\zeta_Y) ~~~~~~~~~
e^{g(x)} = \left\{
\begin{array}{cl} \displaystyle
\left(\frac{\beta}{\cosh((p_X-1)\beta(x-x_0))}
\right)^\frac{1}{p_X-1}  & ~~ \eta_X\sigma_X = -1 \\
\rule{0mm}{6mm}  e^{\pm \beta (x-x_0)} & ~~~ \,~\sigma_X = 0
\rule{0mm}{10mm} \\
 \displaystyle
\left( \frac{\beta}{\sinh((p_X-1)\beta(x-x_0))}
\right)^\frac{1}{p_X-1}  & ~~~ \eta_X\sigma_X = 1
\rule{0mm}{10mm}
\end{array} \right. \label{lage}
\end{equation}
In these expressions, we have introduced two new parameters
$\zeta_Y$ and $\gamma_Y$,
\begin{eqnarray}
\zeta_Y &=& {\rm sign}(\tilde c_Y \gamma_Y)
= \eta_T\eta_Y {\rm sign} (c_Y\gamma_Y) \nonumber\\
\gamma_Y &=&(1+q)c^Y_C + (1+p_Y)c^Y_V - \med \epsilon_Y a_Y
c_\phi^Y ~.
\end{eqnarray}
Notice that $\gamma_Y$ will turn out to be related to the $RR$
charge (the analog of $Q_Y$ in the previous ansatz). We will make
the choice of $c_Y$ such that ${\rm sign} (c_Y\gamma_Y) = +1$  and
therefore everywhere we may substitute $\eta_T\eta_Y$ in place of
~$\zeta_Y$. In the above solution, the independent parameters for
the $x$ dependence can be taken to be $c^Y_C, c^Y_V,c^Y_W,
c^Y_\phi$ and $x_0$. In the context of ordinary non BPS branes
with $c^Y_C = c^Y_V$ (hence full $ISO(1+q+p_Y)$) and $c^Y_W = 0$,
this would be a three parameter family. Also the asymptotic value
of the dilaton has been selected to absorb an additional constant
${c^Y_\phi}'$, and $x_Y$ can be fixed at will by an appropriate
choice of the origin for $x$. We have also made particular scale
transformation on the coordinates to absorb additional constants
(see Appendix B). The rest of the parameters appearing in
(\ref{sys}) and (\ref{lage}) are derived from the previous ones:
\begin{eqnarray}
c^Y_U &=& c^Y_X =   -\frac{1}{p_X-1}\left[ (1+q)c^Y_C +
(1+p_Y)c^Y_V + p_t c^Y_W \right] \\ (\beta^Y)^2  &=&
\frac{1}{p_X(p_X-1)}\left[\frac{1}{p_X-1}
\left(\rule{0mm}{5mm}(1+q)c^Y_C + (p_Y+1)c^Y_V + p_t
c^Y_W\right)^2\right.
\nonumber\\
&&\left. ~~~+ ~(1+q)\left(c^Y_C\right)^2 +
(p_Y+1)\left(c^Y_V\right)^2 + p_t \left(c^Y_W\right)^2 + \med
\left(c^Y_\phi\right)^2\right] \label{bet}
\end{eqnarray}
Actually, these two equations replace (\ref{kki}) and (\ref{ghju})
in the previous ansatz, where now $\beta_Y$ plays the r\^ole of
the non-extremality parameter.

From the off diagonal Einstein equation we obtain the same
intersection rule as before (\ref{intr}) and algebraic constraints
(\ref{xyu}) and (\ref{xyd}), which are sufficient to ensure the
validity of the separability ansatz (\ref{spu}) and (\ref{spd}).
Instead of (\ref{itsc}), in the present setting we get
\begin{eqnarray}
&(p_Y-1) \mu_X (c^Y_C + c^Y_V) + (1+q) c^X_C c^Y_C + (p_Y-1) c^X_V
c^Y_V + (p_X-1) c^X_U c^Y_U & \nonumber\\ &
~~~~~~~~~~~~~~~~~~~~~~~ + p_t c^X_W c^Y_W + 2 c^X_U c^Y_V + \med
c^X_\phi c^Y_\phi = 0 & \label{gpok}
\end{eqnarray}
In the present case, however, we do not have the analog of
(\ref{gud}). This was simply a restriction imposed in order to be
able to obtain harmonic solutions in the  black-brane ansatz. This
is not needed here. On the other side, the $X$ brane constants
$c^X_\alpha$ are still bound by constraints (\ref{nks}),
(\ref{ghjd}) and (\ref{gun}). In total there is one algebraic
equation less and, therefore, we do not need to set one of the
branes to extremality, nor we expect to find discrete sets of
solutions like the ones listed in  Appendix C.

The final expressions for the gauge potentials are
\begin{eqnarray}
E^X &=& E^X_x E^X_y =  ~ e^{\kappa_Y x}
K_x^{\frac{2(D-2)}{\Delta_Y}}~ \sqrt{\eta_T\eta_X\,\frac{Q_X
}{(Q_X+2\mu_X) } \frac{2(D-2)}{ \Delta_X} }     ~
\frac{f_y}{H_y}
\label{eydey}\\
E^Y&=& E^Y_x E^Y_y = ~-\eta_T\eta_Y\sqrt{\frac{2(D-2)}{\Delta_Y}}
~\frac{e^{2\gamma_Y(x-x_Y)}}{K_x} ~~y^{p_Y}f_y^{\chi_X}
H_y^{\frac{2(D-2)}{\Delta_X}} \label{eydex}
\end{eqnarray}
with $\kappa_Y =-\frac{2}{p_X-1}\left(p_Xp_t c^Y_W +
(p_Xp_Y+1)c^Y_V + (1+q) c^Y_C\right) + \epsilon_X a_Xc^Y_\phi$.
The corresponding charge densities (per unit $W$ space volume) are
given by
\begin{eqnarray}
{\cal Q}_X &=& \eta_T\eta_X\eta_Y(p_Y-1)\sqrt{\eta_T\eta_X
Q_X(2\mu_X+Q_X)\frac{2(D-2)}{\Delta_X}}~ \Omega_{p_Y} \nonumber\\
{\cal Q}_Y &=& \eta_T\eta_X\eta_Y\, \sqrt{
\frac{2(D-2)}{\Delta_Y}}~\gamma_Y \,\Omega_{p_X}
\end{eqnarray}
whence we see that $\gamma_Y$ is directly related to the charge.
Contrary to the results in the previous section, notice that there
is no obstruction now against a signature flip of the sort $\eta_T
= \eta_X = -1$, which will be the subject of study in what
follows.

\section{Intersecting $D$-branes and $S$-branes}

In this section, the results obtained in the previous one will be
exploited to generate intersection among supergravity $D$-branes
and $S$-branes through analytic continuation\footnote{Notice that
the name $D$-brane is being used for any extremal object. This is
clearly an abuse of terminology, since we are also considering
configurations involving $NS$ and $M$-theory branes}. To fix the
conventions, we will take $X$ for the $D$-brane, and $Y$ as the
$S$-brane. The solution can be obtained from the previous section
exchanging the roles of $t$ and $x$ as timelike coordinates, by
setting $\eta_T = \eta_X = -1$ and keeping $\eta_Y= +1$. Recall
that the first ansatz applied to the $D$-brane $X$, ({\em i.e.} to
the $y$ dependence) asks for $\sigma_Y = +1$. Moreover the
usefulness of the second ansatz, as applied to the $x$ dependence,
becomes apparent. Indeed, starting from (\ref{seca}) and setting
the black-brane $X$ to criticality \footnote{We might also treat
within our general framework, intersections involving
$D\!-\!\bar{D}$ systems or black-branes with $S$-branes. The
actual solutions can be easily obtained, following the lines of
this section, by borrowing results from sections 3 and 4 of this
paper.} $f_y \to 1$, we may rename $t\to z_{q+1}$ as well as $x
\to t$ and $y \to r$ and thus end up with a line element of the
form
\begin{eqnarray}
ds^2 &=& C^2 d\vec z_p^{\,2} - X^2 dt^2 + U^2
d\Omega^2_{p_D,\sigma_D} + V^2( dr^2 + r^2 d\Omega_{p_S,+1}^2) +
W^2 d \vec w^2_{\,p_t} \label{secag}
\end{eqnarray}
where, in order to stress that $X$ is the $D$-brane and $Y$ the
$S$-brane we have changed labels $X\to D$ and $Y\to S$. Now $p =
q+1$ is the dimension of the Euclidean intersection manifold. Just
to better grasp where the $S$-branes and the $D$-branes are one
may momentarily set $X_r = C_r$ and $V_t = C_t$, which is a
consistent truncation that only amounts to tuning the integration
constants to $c^D_U = c^D_C$, as well as $c^S_V = c^S_C$. To
recover the $D$-brane just skip all dependencies along its
worldvolume coordinate $t$ (and remember that $U_r = X_r$ and $~
V_t = Y_t$ from (\ref{dkmu}))
\begin{equation}
ds^2_D = C^2_r\left( - dt^2 +  d\vec z_p^{\,2} +
d\Omega^2_{p_D,\sigma_D}\right) + V^2_r( dr^2 + r^2
d\Omega_{p_S,+1}^2) + W^2_r d \vec w^{\,2}_{p_t}
\end{equation}
For $\sigma_D=0$ this is just the standard solution for a
$Dq_D$-brane with $q_D = (1+q+p_D)$, which is smeared along a
$p_t$ dimensional outer space. The remnant of the presence of the
intersecting $S$ brane is imprinted in the breaking of the
$SO(1,p+p_D)$ symmetry, which allows us to regard other
possibilities for the curvature $\sigma_D$ other than $+1$.
Conversely, to recover an $S(q_S = p+p_S)$-brane, we shall tune
all functions  ${\cal F}_r \to  1$, thus obtaining the following
line element:
\begin{equation}
ds^2_S =  -X_t^2 dt^2  + C_t^2 (d\vec z_{p}^{\,2} + dr^2 + r^2
d\Omega_{p_S,+1}^2) + U_t^2   d\Omega^2_{p_D,\sigma_D} + W^2_t d
\vec w^{\,2}_{p_t} \label{kwo}
\end{equation}
One may go, of course, to Euclidean coordinates $dz_1,...,
dz_{1+q_S}$ if desired. In accordance with the standard convention
for a $Sq_S$-brane \cite{GS,CGG}, (\ref{kwo}) contains a $q_S+1$
dimensional isotropic ``Euclidean worldvolume".  For completeness,
we reproduce here the full solution for the ansatz (\ref{secag}),
leaving the details of its obtention to Appendix B,
\begin{eqnarray}
X &=&  ~ e^{c_Ut}K_t^{\frac{q_S+1}{\Delta_S}} e^{p_Dg(t)} ~
    H_r^{-\frac{d-q_D-3}{\Delta_D}} \nonumber \\
C &=& ~ e^{c_Ct}K_t^{-\frac{d-q_S-3}{\Delta_S}} ~
    H_r^{-\frac{d-q_D-3}{\Delta_D}}   \nonumber \\
U &=&  ~ e^{c_Ut}K_t^{\frac{q_S+1}{\Delta_S}}  e^{g(t)} ~
    H_r^{-\frac{d-q_D-3}{\Delta_D}}   \label{sol} \\
V &=&   ~ e^{c_Vt}K_t^{-\frac{d-q_S-3}{\Delta_S}} ~
    H_r^{ \frac{q_D+1}{\Delta_D}}  \nonumber \\
W &=&   ~ e^{c_Wt}K_t^{\frac{q_S+1}{\Delta_S}} ~
    H_r^{ \frac{q_D+1}{\Delta_D}}   \nonumber \\
e^{\phi} &=&  ~ e^{c_\phi t}K_t^{\frac{\epsilon_S
a_S(d-2)}{\Delta_S}} ~
    H_r^{ \frac{\epsilon_D a_D(d-2)}{\Delta_A}}  \nonumber \\
&& \nonumber\\
K_t &=& \med(e^{2\gamma_S (t-t_S)}+1)~~~~~~;~~~~ H_r = 1 +
\frac{Q_D}{r^{p_D-1}}
\end{eqnarray}
where we use $d$ for the dimensionality of spacetime to avoid
confusion with the $D$-brane label. On the other hand, the fluxes
are given by
\begin{eqnarray}
E^D &=&  e^{\kappa_D t}K_t^{\frac{2(d-2)}{\Delta_S}} ~\sqrt{
\frac{2(d-2)}{\Delta_D} }
~\left(\frac{1}{H_r}\right) \nonumber\\
E^Y &=&
\sqrt{\frac{2(d-2)}{\Delta_S}}\frac{e^{2\gamma_S(t-t_S)}}{K_t}~
r^{p_S}H_r^{\frac{2(d-2)}{\Delta_D}}
\end{eqnarray}
The integration constants satisfy the same constraints as in the
previous section. Also the intersection rule can be inferred from
(\ref{intr}). It is preferable to re-express it directly in terms
of the data that define the intersection, namely, the dimensions
$q_D$ and $q_S$  of the intervening branes and $p$ of the
intersection manifold
\begin{equation}
p+2 ~ = ~ \frac{(q_D+1)(q_S+1)}{d-2} - \frac{1}{2} \epsilon_D
\epsilon_S a_D a_S \label{intds}
\end{equation}
The set of solutions is summarized in the following table
\begin{eqnarray}
(2|M5, SM5|1)_{11}&& \nonumber\\
(2|NS5,SNS5|0)_{10} &&\nonumber\\
(q_D-2|Dq_D, SN\!S 5|1)_{10}&~~~&   \nonumber\\
(q_S-2| NS 5, SDq_S|1)_{10} &~~~&  \label{irul}\\
(\frac{q_D+q_S}{2}-3|Dq_D,SDq_S|5-\frac{q_D+q_S}{2})_{10} &~~~~&6\leq
q_D+q_S\leq 10  \nonumber
\end{eqnarray}
in the notation $(p| Dq_D,Sq_S|p_t)_{d}$. One should not forget
the limitation in the total number of dimensions, given by $d =
1+p+p_D+(1+p_S) + p_t= 2+q_D+q_S-p+p_t$.

One important reason to think of (\ref{intds})--(\ref{irul}) as a
new intersection rule is that it encompasses the interesting case
$p=0$ which cannot be obtained by an analytic continuation from a
semi-local timelike intersection (where it would force us to start
with $q=-1$). This case has to be solved separately from scratch,
and the final answer is given by the above solution after setting
$p=0$. The set of solutions presented in (\ref{sol})--(\ref{irul})
provides a supergravity description of intersecting $D$-branes and
$S$-branes.

\section{Discussion and Prospects}

In this paper we have explored a family of supergravity solutions
with the correct isometries to represent pairwise partially
localized intersections of any pair of the following objects:
$D$-branes, $E$-branes, black-branes, $S$-branes, $D\!-\!\bar{D}$
systems and Schwarzschild branes. We have developed in full detail
the cases corresponding to intersection of black-branes with
black-branes or a $D\!-\!\bar{D}$ system, and that of $S$-branes
intersecting $D$-branes. These solutions are of the factorized
type, a key assumption for achieving the separation of the
equations of motion. We have put to zero the separation constants,
and this is certainly a particular case that makes the integration
feasible. Certainly an analysis of solutions where this is not so
is a straightforward extension of the present work.

After rotating one radial coordinate and the time, each of these
intersection yields p-brane solution with a time dependence of the
$S$-brane type. Applications to cosmology should proceed as
follows. First of all, identify among the factor spaces, a three
dimensional maximally  symmetric one lying inside the $D$-brane,
and compactify all the rest. After going over to the Einstein
frame, a specific four dimensional cosmological evolution emerges
automatically. In this respect, let us mention that time dependent
compactifications involving direct product of several factor
spaces may lead to inflation with a sufficient number of
e-foldings \cite{CHNW} (contrary to what happens whenever there is
a single outer space \cite{EG}, where a transient period of
accelerated expansion is achieved but the number of e-foldings is
around $1$; see also \cite{Russo,NeuV} for related results and
generalizations). In our framework, however, the way these factor
spaces appear is tightly constrained by the intersection rules and
the need of an ISO(3) symmetric subspace. A detailed analysis of
the cosmological features of these solutions will appear soon
\cite{EM}

Another avenue for future research has to do with the possibility
to formulate AdS/CFT duality in a time dependent context. In order
to do so, the near horizon geometries have to be carefully
unravelled and, for that, a regular supergravity $S$-brane
solution is probably needed. The method presented in this paper
attempts to be a first step in that direction. It would be
interesting to examine the very recent regular $S$-brane solutions
found in \cite{Adv} within this context. Of course, issues like
the near horizon limit of such time dependent configurations are
highly non-trivial. It would also be interesting to see whether
the addition of tachyonic matter changes the singularity patterns
of the configurations discussed in this paper or not. In this
respect, we would like to point out that the most natural
configuration to study these issues displays $D3$-branes
intersecting space-filling neutral $SD8$-branes, and it is
precisely for the latter that singularity theorems seem to be
elusive \cite{BLW,BW}.

The microscopic description of the intersections of $D$-branes and
$S$-branes might shed light into many of these issues and is worth
studying.

~

{\em Note added:} Soon after submission of this manuscript,
configurations involving S-branes and (space filling) D-branes
have also been considered in studying a possible mechanism to
resolve the aforementioned S-brane supergravity solutions'
singularities \cite{Leb}.

\section*{Acknowledgments}

We would like to express our gratitude to Max Ba\~nados, Roberto
Emparan, Martin Kruczenski, Alfonso Ramallo, Jorge Russo, Kostas
Skenderis and Stefan Theisen for helpful comments and/or
correspondence. JDE and JM are pleased to thank respectively the
Theory Group in Santiago de Compostela and the Physics Laboratory
at CECS for most warm hospitality while this work was being done.
We both wish to thank the Benasque Center for Science where it was
initiated.

This work has been supported in part by MCyT, FEDER and Xunta de
Galicia under grant BFM2002-03881, by Fundaci\'on Antorchas, by
the FCT (Portugal) grant POCTI/FNU/38004/2001, and by the EC
Commission under the FP5 grant HPRN-CT-2002-00325. Institutional
support to the Centro de Estudios Cient\'\i ficos (CECS) from
Empresas CMPC is gratefully acknowledged. CECS is a Millennium
Science Institute and is funded in part by grants from Fundaci\'on
Andes and the Tinker Foundation.

\appendix

\section{Solving for Intersecting Black-Branes}

In this appendix we shall give the details of the general solution
used in the text. The procedure follows the constructive  approach
of \cite{Ohta-ne} but we try to keep as general as possible. We
shall integrate the dependence on one of the coordinates $x$. It
should be noted that, regarding the $x$ radial coordinate living
inside the $X$ brane, all the dependence on it comes from the
presence of the $Y$ brane. Therefore the integration constants
carry an index $Y$. The analog equations for $y$ dependence can be
obtained after replacing $x \to y$, $X \to Y$ and $U \to V$. After
inserting the gauge fixing ansatz (\ref{anu}) into (\ref{eqse})
all the equations except the ones for $X_x$  may be integrated
once to give
\begin{equation}
\begin{array}{rcl}
f_x(\ln T_x)' &=& \displaystyle \frac{d^Y_{||}
    E^Y_x + c^Y_T}{x^{p_X}} \\
f_x(\ln C_x)' &=\rule{0mm}{10mm}&\displaystyle \frac{d^Y_{||}
    E^Y_x + c^Y_C}{x^{p_X}}  \\
f_x(\ln xU_x)' &=\rule{0mm}{10mm}&\displaystyle \frac{d^Y_\perp
    E^Y_x + c^Y_U}{x^{p_X}} + \frac{\eta_X\sigma_X}{x} \\
\end{array}
~~~~~~~~~
\begin{array}{rcl}
f_x(\ln V_x)'  &=\rule{0mm}{10mm}&\displaystyle \frac{d^Y_{||}
    E^Y_x + c^Y_V}{x^{p_X}}   \\
f_x(\ln W_x)' &=\rule{0mm}{10mm}&\displaystyle \frac{d^Y_{\perp}
    E^Y_x + c^Y_W}{x^{p_X}}  \\
f_x\phi_x' &=\rule{0mm}{10mm}&\displaystyle \frac{-e^Y E^Y_x +
    c^Y_\phi}{x^{p_X}}  \\
\end{array} \label{smi}
\end{equation}
Substituting $T_x = C_xf^{1/2}_x$ and subtracting from the
equation for $C_x$, the following solution for $f_x$ is readily
found
\begin{equation}
f_x = 1-\frac{2\mu_Y}{x^{p_X-1}} \label{efex}
\end{equation}
with $c^Y_T = \mu_Y(p_X-1) + c^Y_C$. Notice, however, that it is
also possible to find a differential equation for $f_x$ by adding
up the following Einstein equations: $(q+1) R^i{_i} + (p_X-1)
R^{\theta_x}{_{\theta_x}} + (p_Y+1) R^{\theta_y}{_{\theta_y}} +
p_t R^w{_w}$,
\begin{equation}
(p_X-1) (f_x x^{p_X-1})' - \eta_X\sigma_X (p_X-1)^2 x^{p_X-2}=0
\end{equation}
since the right hand side adds up to $((q+1+p_Y+1)d^Y_{||} +
(p_X-1-p_t)d^Y_{\perp})E'= 0$. The solution of this equation, with
the right asymptotics, $f_x(\infty)\to 1$, is (\ref{efex}), with the
additional constraint
\begin{equation}
\sigma_X \eta_X = 1 \label{etsi}
\end{equation}
The same reasoning, of course, applies to the Y brane. The gauge
fixing constraints (\ref{gauu}) (\ref{gaud}) impose the following
equality
\begin{equation}
(q+1)c^Y_C  +(p_X-1)(2\mu_Y+c^Y_U) +  (p_Y+1) c^Y_V + p_t c^Y_W =
0 \label{kki2}
\end{equation}
with an analog expression for the other brane. In order to achieve
the full integration we cast the equation (\ref{eqse}) for $X_x$
in the following form
\begin{eqnarray}
&&f_x(f_x(\ln xU_x)')' - (f_x'+\frac{p_X-2}{x}f_x)(f_x(\ln xU_x)')
+ \med f_xf_x'' + \med(\frac{p_X}{x} f_x -\med f_x')f_x'\nonumber\\
&&~+ (f_x\ln T_x)'{^2} + q(f_x\ln C_x'){^2} + (p_X-1) (f_x(\ln
xU_x)'){^2} + (p_Y+1)(f_x \ln V_x'){^2} + p_t(f_x\ln W_x'){^2}
\nonumber\\ && ~~~~~=  ~-\med (f_x\phi_x'){^2} + d_{||}^Y
\frac{f_x{E^Y_x}'}{x^{p_X}}
\end{eqnarray}
Inserting the first integrals, the term which is independent of
$E^Y_x$ yields the following algebraic constraint
\begin{equation}
(p_X-1) c^{Y2}_U+ 2\mu_Y (p_X-1)(c^Y_U+c^Y_C) +  (q+1) c^{Y2}_C +
(p_Y+1)c^{Y2}_V + p_t c^{Y2}_W + \med c^{Y2}_\phi ~=~0 \label{qad}
\end{equation}
The $E^Y_x$ dependent terms, on the other hand, yield the
following equation
\begin{equation}
\left[
\left(\frac{f_x}{E^Y_x}\right)'-\frac{\Gamma_Y}{x^{p_X}E^Y_x}
+\frac{1}{x^{p_X}}\left(\frac{\tilde c_Y}{2}\right)
\frac{\Delta_Y}{(D-2)} \right] {E^Y_x}^2 = 0
\end{equation}
 with
\begin{eqnarray}
\displaystyle \Gamma^Y &=&
2(p_X-1)(2\mu_Y+c^Y_U)+2p_tc^Y_W+\epsilon_Y a_Yc^Y_\phi
\nonumber\\
\Delta_Y &=& (q_Y+1)(D-q_Y-3)+ \med  (D-2) a_Y^2 \label{lad}
\end{eqnarray}
Analytic integration may be easily achieved in the particular case
that we set $\Gamma_Y = 0$. This is really a reduction by one in
the number of parameters the solution may depend on. It would be
interesting to lift this constraint in order to solve for the most
general case. Now, for  $\Gamma_Y= 0$, the resulting equation for
$E^Y_x$ can be solved as
\begin{equation}
E^Y_x = \left(\frac{Q_Y (p_X-1)}{\tilde
c_Y}\frac{2(D-2)}{\Delta_Y}\right) \frac{f_x}{H_x} \label{gaup}
\end{equation}
in terms of a harmonic function $H_x$ in relative transverse space
along the $X$ brane
$$
H_x = 1 + \frac{Q_Y}{x^{p_X-1}}
$$
With these ingredients we can readily obtain the integrated
expressions
\begin{eqnarray}
C_x &=&  f_x^{\frac{c^Y_C}{2\mu_Y(p_X-1)}}
    H_x^{-\frac{D-q_Y-3}{\Delta_Y}} \nonumber\\
X_x &=& U_x f_x^{-1/2} \rule{0mm}{7mm} \nonumber\\
U_x &=& f_x^{\frac{c^Y_U+2\mu_Y}{2\mu_Y(p_X-1)}}
    H_x^{\frac{q_Y+1}{\Delta_Y}} \nonumber\\
Y_x=V_x &=& f_x^{\frac{c^Y_V}{2\mu_Y(p_X-1)}}
    H_x^{-\frac{D-q_Y-3}{\Delta_Y}} \nonumber\\
W_x &=& f_x^{\frac{c^Y_W}{2\mu_Y(p_X-1)}}
    H_x^{\frac{q_Y+1}{\Delta_Y}} \nonumber\\
e^{\phi_x} &=& f_x^{\frac{c^Y_\phi}{2\mu_Y(p_X-1)}}
    H_x^{\frac{\epsilon_Ya_Y(D-2)}{\Delta_Y}} \nonumber\\
\end{eqnarray}
which are enough to compute the functions $S^A_x$. Using
(\ref{maxs}) we obtain
\begin{equation}
E^X_x  = \frac{l_X}{S^X_x} = l_X\, x^{p_X}f_x^{\chi_Y}
H_x^{\frac{2(D-2)}{\Delta_Y}} ~~~
\end{equation}
with
\begin{equation}
\chi_Y =  \frac{-1}{\mu_Y(p_X-1)}\left(p_t c^Y_W + p_Y c^Y_V -
(c^Y_U+2\mu_Y) + \med \epsilon_X a_X c^Y_\phi\right) \label{chiy}
\end{equation}
Also, we may compute $S^Y_x(\d_x E^Y_x)$ and equate it to $c_Y$ as
in (\ref{maxs}). Using (\ref{thj})  is enough to solve for the
product $c_Yl_Y$ as
\begin{equation}
c_Yl_Y = (p_X-1)\sqrt{\eta_T\eta_Y
Q_Y(2\mu_Y+Q_Y)\frac{2(D-2)}{\Delta_Y}} \label{dki}
\end{equation}
This can be inserted back into (\ref{gaup}) to find
\begin{equation}
E^Y_x  = \frac{\eta_T\eta_Y}{l_Y}\sqrt{\eta_T\eta_Y\,
\frac{Q_Y}{(2\mu_Y+Q_Y)}\frac{2(D-2)}{\Delta_Y}} ~\frac{f_x}{H_x}
\end{equation}
Concerning the off diagonal Einstein equation, writing down $f
x^{p_X} y^{p_Y} R_{xy}$ explicitly, we may separate the
coefficients of different powers of $E^Y_x$ and $E^X_y$. The terms
proportional to $E^XE^Y$ yield the same intersection rule as in
the extremal case
\begin{equation}
q+3 = \frac{(q_X+1)(q_Y+1)}{(D-2)}
-\frac{1}{2}\epsilon_X\epsilon_Y a_X a_Y \label{rkj}
\end{equation}
From the terms proportional to $E^Y$ and $E^X$ we obtain the
following constraints
\begin{eqnarray}
&&p_t c^Y_W + (p_Y-1) c^Y_V + \med \epsilon_X a_X c^Y_\phi = 0
\label{xyuv}
\\ &&p_t c^X_W + (p_X-1) c^X_U + \med \epsilon_Y a_Y c^X_\phi = 0
\label{xyde}
\end{eqnarray}
and finally from the independent terms
\begin{eqnarray}
&&\mu_X \mu_Y (p_X-1) (p_Y-1) + \mu_X (p_Y-1) (c^Y_V+c^Y_C) +
\mu_Y (p_X-1) (c^X_U+c^X_C) + (q+1) c^X_C c^Y_C \nonumber\\ &&
~~~~~~~ + (p_X-1)c^X_Uc^Y_U + (p_Y-1)c^X_Vc^Y_V +p_t c^X_W c^Y_W +
2 c^X_U c^Y_V + \med c^X_\phi c^Y_\phi =0 \rule{0mm}{7mm}
\end{eqnarray}
One can easily check that these equations are enough to verify the
separability assumption (\ref{spu}) and (\ref{spd}).

The case with $p_X=1$ can also be worked along the same lines. In
this case, the $Y$ brane is a co-dimension 2 defect inside the $X$
brane and therefore the harmonic functions involve logarithms of
$x$. It is always understood that a completely analogous analysis
goes through for the $y$ dependence.

\section{Solving for Black-branes intersecting $D\!-\!\bar{D}$ system}

For the second $D\!-\!\bar{D}$ system we will solve again only for
the $x$ dependence, caused as before by the $Y$ brane. Also here,
we shall follow a generalization of the approach in \cite{Ohtair}.
On top of $T = C$, the gauge fixing constraint will be taken to be
$f_x x^{p_X} = 1$, or equivalently
\begin{equation}
1 =  C^p_x U^{p_X}_x V^{p_Y+1}_x W^{p_t}_x X^{-1}_x \label{gaut}
\end{equation}
Now $p = q+1$ and $q_A = p + p_A$ is the dimension of each brane.
The $x$ system can be first-integrated, except for the equations
involving $X$ and $U$
 \begin{eqnarray}
&& (\ln C_x)' =   d^Y_{||} E^Y_x + c^Y_C \nonumber\\
&& (\ln X_x)'' -{(\ln X_x)'}^2  + p{(\ln C_x)'}^2 +p_X{(\ln U_x)'}^2
+ (p_Y+1){(\ln V_x)'}^2 + p_t {(\ln W_x)'}^2 \nonumber\\
&&~~~~~~~~~~~~~~~~~~~~~~~~ = -~\frac{1}{2}{\phi_x'}^2 + d^Y_{||}
{E^Y_x}' \label{lax}\\ && (\ln U_x)''  -
\eta_X\sigma_X(p_X-1)\left(\frac{X_x}{U_x}\right)^2 =
d^Y_{\perp}{E^Y_x}'\label{lau}\\
&& (\ln V_x)'   =   d^Y_{||} E^Y_x + c^Y_V\nonumber\\
&& (\ln W_x)'  =   d^Y_{\perp} E^Y_x + c^Y_W\nonumber\\
&& \phi_x'  = -e^YE^Y_x + c^Y_\phi
\end{eqnarray}
Let us define, for conciseness,
\begin{eqnarray}
\sum_\alpha \ln U_{x\,\alpha} &=& p\ln C_x + (p_Y+1)\ln V_x + p_t
\ln W_x \nonumber\\
\sum_\alpha c^Y_\alpha &=& pc^Y_C + (p_Y+1)c^Y_V + p_t c^Y_W\\
\sum_{\alpha||Y} c^Y_\alpha &=& pc^Y_C + (p_Y+1)c^Y_V \nonumber
\end{eqnarray}
The non-extremality resides now in the distinction among $\ln X$
and $\ln U$. Following \cite{GS,CGG,Ohtair}, we will parameterize
this through a new function
\begin{equation}
g = \frac{\ln X_x - \ln U_x}{p_X-1}
\end{equation}
Recalling the constraint (\ref{gaut}) it follows that
\begin{eqnarray}
\ln U_x &=& g - \frac{1}{p_X-1} \sum_\alpha \ln U_{x\,\alpha}
\nonumber\\ \ln X_x &=& p_X g - \frac{1}{p_X-1} \sum_\alpha \ln
U_{x\,\alpha}
\end{eqnarray}
and taking derivatives
\begin{eqnarray}
(\ln U_x)' &=& g' + d^Y_\perp E^Y_x ~ +~ c^Y_U  \\
(\ln X_x)' &=& p_X g' + d^Y_\perp E^Y_x ~+~ c^Y_X
\end{eqnarray}
where $c_X = c_U = - \frac{1}{p_X-1} \sum_\alpha c_\alpha.$ After
inserting these expressions into equation (\ref{lau}), a
differential equation for $g$ is found
\begin{equation}
g'' - \eta_X\sigma_X(p_X-1) e^{2(p_X-1)g} = 0
\end{equation}
with the solution
$$
g(x) = \left\{
\begin{array}{cl} \displaystyle
\frac{1}{p_X-1} \ln\frac{\beta}{\cosh((p_X-1)\beta(x-x_0))} &
~~~~~~\eta_X\sigma_X = -1 \\ \rule{0mm}{6mm}  \pm \beta (x-x_0) &
~~~~~~\sigma_X = 0 \\ \displaystyle  \frac{1}{p_X-1}
\ln\frac{\beta}{\sinh((p_X-1)\beta(x-x_0))}  &
~~~~~~\eta_X\sigma_X = 1
\end{array}
\right.
$$
From this result it is clear that $\beta$ plays the r\^ole of the
non-extremality parameter. Indeed the limit $\beta \to 0$ is well
defined only for the cases $\sigma_X = 0$ and $\eta_X\sigma_X =
1$, leading to a flat relatively transverse $p_X+1$ dimensional
space.

Noticing that $g'' - (p_X-1) (g')^2 = -(p_X-1)\beta^2$ we may
insert this into equation (\ref{lax}) and obtain
\begin{eqnarray}
&&-p_X(p_X-1)\beta^2 + (d^Y_\perp - d^Y_{||}) {E^Y_x}' + (p_X-1)
\left (d^Y_\perp E^Y_x - \sum_\alpha \frac{c^Y_\alpha}{p_X-1}
\right)^2 + p (d^Y_{||} E^Y_x + c^Y_C)^2 \nonumber\\ && ~~~~~~~~~
+ (p_Y+1) (d^Y_{||} E^Y_x + c^Y_V)^2 + p_t (d^Y_\perp E^Y_x +
c^Y_W)^2 + \med( -e^Y E^Y_x + c^Y_\phi)^2 = 0
\end{eqnarray}
From the constant term we find
\begin{equation}
\frac{1}{p_X-1}\left( \sum_\alpha c^Y_\alpha \right)^2 +
\sum_\alpha {c^Y_\alpha}^2  + \med {c^Y_\phi}^2 =
p_X(p_X-1)\beta^2 \label{dlkjl}
\end{equation}
and from the $E^Y_x$ dependent part,
\begin{equation}
\left(\frac{1}{E^Y_x}\right)'+ \frac{2\gamma_Y}{E^Y_x} =
-\frac{\Delta_Y\tilde c_Y}{2(D-2)} \label{krs}
\end{equation}
where $\Delta_Y$ is the same as in (\ref{lad}) and
\begin{equation}
\gamma_Y = \sum_{\alpha|| Y} c^Y_\alpha - \med a_Yc^Y_\phi
\end{equation}
Equation (\ref{krs}) can be readily integrated as follows
\begin{eqnarray}
E^Y_x &=& -\frac{2(D-2)}{\Delta_Y} \frac{\gamma_Y}{\tilde c_Y}
~\frac{e^{2\gamma_Y(x-x_Y)}}{K_x} \label{eyx} .
\end{eqnarray}
with
$$
K_x = \frac{1}{2} (e^{2\gamma_Y(x-x_Y)}-\zeta_Y)
~~~~~~~~;~~~~~~~~~\zeta_Y ={\rm sign}({\tilde c_Y \gamma_Y})
$$
It is now straightforward to  write down the complete solution
\begin{eqnarray}
C_x &=& e^{c^Y_C x + {c^Y_C}' }(e^{\gamma_Y
x_Y}K_x)^{-\frac{D-q_Y-3}{\Delta_Y}} \nonumber \\
X_x &=& e^{ c^Y_X x + {c^Y_X}'} e^{p_Xg(x)}
    (e^{\gamma_Y x_Y}K_x)^{\frac{q_Y+1}{\Delta_Y}} \nonumber \\
U_x &=& e^{ c^Y_U x +{c^Y_U}'} e^{g(x)}
    (e^{\gamma_Y x_Y}K_x)^{\frac{q_Y+1}{\Delta_Y}} \nonumber \\
Y_x = V_x &=& e^{c^Y_V x+{c^Y_V}'}
    (e^{\gamma_Y x_Y}K_x)^{-\frac{D-q_Y-3}{\Delta_Y}} \nonumber \\
W_x &=& e^{c^Y_W x+{c^Y_W}' }
    (e^{\gamma_Y x_Y}K_x)^{\frac{q_Y+1}{\Delta_Y}} \nonumber \\
e^{\phi_x} &=& e^{c^Y_\phi x+{c^Y_\phi}'}
    (e^{\gamma_Y x_Y}K_x)^{\frac{(D-2)a_Y\epsilon_Y}{\Delta_Y}}
\end{eqnarray}
where ${c^Y_X}' = {c^Y_U}' = \frac{-1}{p_X-1} \sum_\alpha
{c^Y_\alpha}' .$ The constants ${c^Y_\alpha}' = 0$ and
$e^{\gamma_Y x_Y}$ can be absorbed by a rescaling of the
coordinates and we have done so in the main body of the paper and
hereafter. The rescaling of $x$ and $y$ needed forces a
compensating redefinition of all the constants $c^Y_\alpha$ and
$c^X_\alpha$, including $\mu_X$ and $Q_X$. After doing so, an
explicit calculation yields the following results
\begin{eqnarray}
S^Y_x &=&  \frac{W^{p_t}_x U^{p_X}_x}{C^p_x X_x
V^{p_Y+1}_x}e^{a_Y\phi_x}  =  e^{-2\gamma_Yx} K_x^2
\label{bu}\\
S^X_x &=& \frac{W^{2p_t}V^{2p_Y}}{X^2_x} e^{a_X\phi_x}~=~
e^{\kappa_Y x} K_x^{-\frac{2(D-2)}{\Delta_Y}}
\end{eqnarray}
with
$\kappa_Y = -\frac{2}{p_X-1}\left(p_Xp_t c^Y_W + (p_Xp_Y+1)c^Y_V + p
c^Y_C\right) + a_Xc^Y_\phi.
$
These functions are needed in order to solve for $E^X_x$
\begin{equation}
E^X_x = \frac{l_X}{S^X_x} = l_X e^{\kappa_Y x}
K_x^{\frac{2(D-2)}{\Delta_Y}}
\end{equation}
as well as to compute the normalization
\begin{equation}
S^Y_x \d_x E^Y_x  =  {\rm sign}(\tilde c_Y
\gamma_Y)\frac{2(D-2)\gamma_Y^2}{\Delta_Y\tilde c_Y} = c_Y
\end{equation}
From here one readily gets the $RR$ charge of the $Y$ brane
\begin{equation}
{\cal Q}_Y = \eta l_Y c_Y \Omega_{p_Y}  =  \eta \sqrt{{\rm sign}(
c_Y \gamma_Y)\frac{2(D-2)}{\Delta_Y}}~\gamma_Y  \Omega_{p_Y}
\label{cha}
\end{equation}
with $\eta = \eta_T\eta_X\eta_Y$. We see that the parameter
$\gamma_Y$ is related to the charge and plays the r\^ole of $Q_Y$
in the other ansatz.  Since $\zeta _Y = {\rm sign}(\tilde
c_Y\gamma_Y) =\eta_T\eta_Y{\rm sign}(c_Y\gamma_Y)$, after choosing
sign$(c_Y\gamma_Y)= +1$ to ensure reality of the charge
(\ref{cha}), we are allowed to replace $\zeta_Y = \eta_T\eta_Y$ in
all the expressions. Now, for the gauge potential we obtain
\begin{equation}
E^Y_x= ~-\frac{\eta_T\eta_Y}{l_Y}\sqrt{
\frac{2(D-2)}{\Delta_Y}}~\frac{e^{2\gamma_Y(x-x_Y)}}{K_x}
\end{equation}
and we observe that the reality is independent of the signature
choice. For the $y$ dependence we may  find a specular solution
after replacing $x \to y, U\leftrightarrow V$ and
$Y\leftrightarrow X$. Alternatively, we may mix this ansatz with
an ansatz of the form given for a black brane in the previous
section. Then multiplying functions of $x$ with functions of $y$
the full solution is obtained.

\section{Bestiary of solutions}

In this appendix we shall give the explicit solutions for the
intersection of black-branes  $p_X,p_Y\neq 1$. The harmonic
functions are the ones given in (\ref{efes}) and (\ref{aches}).
The parameters $\xi^A,~ A = X,Y.$ stand for
$$
\xi^A = \left(1 + \frac{2\mu_A}{Q_A}\right)^{-1/2} .
$$
First of all, the list of solutions that depend on three
parameters contains the cases $(p-3|Dp,NS5|1)$, $4\leq p\leq 7.$
Their actual expressions are a bit cumbersome so we are not going
to include them, at least explicitly, in the bestiary.

\subsection{Solutions depending on two parameters}

Let us list the only five cases of two parameter (say,
$\mu_X,\mu_Y$) family of solutions. As discussed in the body of
the paper, for all these configurations we have found that one of
the intervening branes is {\em minimally non-extremal}. We use
$d\Omega_k^2$ for the round metric of the unit $k$-sphere.

\subsubsection*{$(1| D2, D8|0)$}
We find two solutions, depending on the intervening brane that is
minimally non-extremal. When it is the $D2$-brane,
\begin{eqnarray}
ds^2 &=& H_x^{9/8}H_y^{3/8} f_x^{-9/8} [H_x^{-1}H_y^{-1} (-f_xf_y
dt^2+ dz^2) + H_y^{-1} f_x^{-3} dx^2 + H_x^{-1} f_x (f^{-1}_y dy^2
+ y^2 d\Omega_6^2)] \nonumber\\
e^\phi &=& f_x^{5/4} H_x^{-5/4} H_y^{1/4}~~ ;~~E^X = \xi^X H_x
f_x^{-3} H_y^{-1} f_y~;~~ E^Y = \xi^Y H_x^{-1} f_x ~y^6 H_y
\end{eqnarray}
whereas for the $D6$-brane, it reads
\begin{eqnarray}
ds^2 &=& H_x^{9/8}H_y^{3/8} f_y^{-3/8} [H_x^{-1}H_y^{-1} (-f_xf_y
dt^2+ dz^2) +H_y^{-1} f_y (f_x^{-1} dx^2) + H_x^{-1}
f_y^{2/5}(f_y^{-1} dy^2 + y^2 d\Omega_6^2)] \nonumber\\
e^\phi &=& f_y^{-1/4} H_x^{-5/4} H_y^{1/4}~;~~E^X = \xi^X H_x f_y
H_y^{-1} ~;~~ E^Y = \xi^Y H_x^{-1} f_x ~y^6 H_y f_y^{-3/5}
\end{eqnarray}

\subsubsection*{$(1| D4, D6|0)$}
Again, we have two solutions according to whether it is the
$D4$-brane or the $D6$-brane the minimally non-extremal one. When
it is the $D4$-brane, we get
\begin{eqnarray}
ds^2 &=& H_x^{7/8}H_y^{5/8} f_x^{-7/8} [H_x^{-1}H_y^{-1} (-f_xf_y
dt^2+ dz^2) + H_y^{-1} f_x^{2} (f^{-1}_x dx^2 + x^2 d\Omega_2^2)
\nonumber\\ && ~~~~~~~~~~~~~ + H_x^{-1} f_x(f_y^{-1} dy^2 + y^2
d\Omega_4^2)] \nonumber\\
e^\phi &=& f_x^{3/4} H_x^{-3/4} H_y^{-1/4}~;~~ E^X = \xi^X  x^2
H_x f_x H_y^{-1} f_y~;~~ E^Y = \xi^Y H_x^{-1} f_x ~y^4 H_y
\end{eqnarray}
whereas for the $D6$-brane, we obtain
\begin{eqnarray}
ds^2 &=& H_x^{7/8}H_y^{5/8} f_y^{-5/8} [H_x^{-1}H_y^{-1} (-f_xf_y
dt^2+ dz^2) + H_y^{-1} f_y (f^{-1}_x dx^2+ x^2 d\Omega_2^2)
\nonumber\\ && ~~~~~~~~~~~~~ + H_x^{-1} f_y^{2/3}(f^{-1}_y dy^2 +
y^2 d\Omega_4^2)] \nonumber\\
e^\phi &=& f_y^{1/4} H_x^{-3/4} H_y^{-1/4}~;~~ E^X = \xi^X  x^2
H_x H_y^{-1} f_y~;~~ E^Y = \xi^Y H_x^{-1} f_x ~y^4 H_y f_y^{-1/3}
\end{eqnarray}

\subsubsection*{$(1| D5, D5|0)$}
We assume that the first $D5$-brane is the minimally non-extremal
one:
\begin{eqnarray}
ds^2 &=& H_x^{3/4}H_y^{3/4} f_x^{-3/4} [H_x^{-1}H_y^{-1} (-f_xf_y
dt^2+ dz^2) + H_y^{-1} f_x (f^{-1}_x dx^2 + x^2 d\Omega_3^2)
\nonumber\\ && + ~~~~~~~~~~~~~ H_x^{-1} f_x(f^{-1}_y dy^2 + y^2
d\Omega_3^2)] \nonumber\\
e^\phi &=& f_x^{1/2} H_x^{-1/2} H_y^{-1/2}~;~~ E^X = \xi^X x^3 H_x
H_y^{-1} f_y~;~~ E^Y = \xi^Y H_x^{-1}f_x ~y^3 H_y
\end{eqnarray}
The solution with $(x\leftrightarrow y)$ and $(X\leftrightarrow
Y)$ just interchanges the r\^ole of the intervening $D5$-branes.

\subsubsection*{$(1| NS5, NS5|0)$}
It is exactly as in the $(1| D5, D5|0)$ solution except for the
dilaton, $e^\phi \to e^{-\phi}$. Up to this point, notice that the
solutions are fully localized.

\subsubsection*{$(1| M5, M5|1)$}
The only solution appearing in eleven dimensional supergravity.
Taking the first $M5$-brane as the minimally non-extremal one, the
solution reads
\begin{eqnarray}
ds^2 &=& H_x^{2/3}H_y^{2/3} f_x^{-2/3} [H_x^{-1}H_y^{-1} (-f_xf_y
dt^2+ dz^2) + H_y^{-1} f_x (f^{-1}_x dx^2 + x^2 d\Omega_3^2)
\nonumber\\ && ~~~~~~~~~~~~~ + H_x^{-1} f_x(f_y^{-1} dy^2 + y^2
d\Omega_3^2) + dw^2] \nonumber\\ E^X &=&  \xi^X x^3 H_x H_y^{-1}
f_y~;~~ E^Y = \xi^Y y^3 H_x^{-1} f_x H_y
\end{eqnarray}
Again, the solution with $(x\leftrightarrow y)$ and
$(X\leftrightarrow Y)$ just interchanges the r\^ole of the
$M5$-branes.

\subsection{Solutions depending on one parameter}

Finally we list the solutions that depend on one parameter. They
necessarily have either $\mu_X=0$ or $\mu_Y=0$, that is, one of
the intervening branes is forced to be extremal.

\subsubsection*{$(0|D1,D7|1)$}
When the $D1$-brane is extremal, $\mu_X = 0$, and the $D7$-brane
can be minimally non-extremal
\begin{eqnarray}
ds^2 &=& H_xH_y^{1/4}[-H_x^{-1}H_y^{-1}f_x dt^2 + H_y^{-1}
f_x^{-1}dx^2 + H_x^{-1} (dy^2 + y^2 d\Omega_6^2) + dw^2] \nonumber\\
e^\phi &=& H_x^{-1} H_y^{1/2}~;~~ E^X  = H_x H_y^{-1}~;~~ E^Y =
\xi^Y H_x^{-1} f_x ~y^6 H_y
\end{eqnarray}
or a non-minimal non-extremal extension
\begin{eqnarray}
ds^2 &=& H_xH_y^{1/4}[-H_x^{-1}H_y^{-1}f_x  dt^2 + H_y^{-1}
f_x^{114/7} dx^2 + f_x^{22/7}H_x^{-1}( dy^2 + y^2 d\Omega_6^2)
+f_x^{-33/7}dw^2] \nonumber\\
e^\phi &=& f_x^{-11} H_x^{-1} H_y^{1/2} ~;~~ E^X  = H_x
f_x^{99/7}H_y^{-1}~;~~ E^Y = \xi^Y H_x^{-1} f_x ~y^6 H_y
\end{eqnarray}
When it is the $D7$-brane the extremal one, $\mu_Y = 0$, the
$D1$-brane might be minimally non-extremal
\begin{eqnarray}
ds^2 &=& H_xH_y^{1/4}[-H_x^{-1}H_y^{-1}f_y dt^2 + H_y^{-1} dx^2 +
H_x^{-1}(f_y dy^2 + y^2 d\Omega_6^2) +dw^2] \nonumber\\
e^\phi &=& H_x^{-1} H_y^{1/2} ~;~~ E^X = \xi^X   H_x H_y^{-1}
f_y~;~~ E^Y = H_x^{-1} ~y^6 H_y
\end{eqnarray}
or a non-minimal non-extremal extension
\begin{eqnarray}
ds^2 &=& H_xH_y^{1/4}[-H_x^{-1}H_y^{-1}f_y dt^2 + f_y^{-2/19}
H_y^{-1} dx^2 + f_y^{8/95}H_x^{-1}(f_y dy^2 + y^2 d\Omega_6^2)
+ f_y^{-6/19}dw^2] \nonumber\\
e^\phi &=& H_x^{-1} H_y^{1/2}f_y^{-2/19} ~;~~ E^X = \xi^X H_x
H_y^{-1} f_y~;~~ E^Y = H_x^{-1} ~y^6 H_y f_y^{18/95}
\end{eqnarray}

\subsubsection*{$(0|D3,D5|1)$}
When the $D3$-brane is extremal, $\mu_X = 0$, the $D5$-brane can
be minimally non-extremal
\begin{eqnarray}
ds^2 &=& H_x^{3/4}H_y^{1/2}[-H_x^{-1}H_y^{-1}f_x dt^2 + H_y^{-1}
(f_x^{-1} dx^2+x^2d\Omega_2^2) + H_x^{-1}(dy^2 + y^2 d\Omega_4^2)
+dw^2] \nonumber\\ e^\phi &=& H_x^{-1/2} ~;~~ E^X =  x^2 H_x
H_y^{-1}~; ~~ E^Y = \xi^Y H_x^{-1} f_x ~y^4 H_y
\end{eqnarray}
or a non-minimal non-extremal extension
\begin{eqnarray}
ds^2 &=& H_x^{3/4}H_y^{1/2}[-H_x^{-1}H_y^{-1}f_x dt^2 + H_y^{-1}
f_x^{2/17}(f_x^{-1}dx^2+x^2 d\Omega_2^2) + H_x^{-1}f_x^{-1/17}
(dy^2 + y^2 d\Omega_4^2)\nonumber\\ && + f_x^{3/17}dw^2] \nonumber\\
e^\phi &=& H_x^{-1/2} f_x^{5/17} ~;~~ E^X =  x^2 H_x f_x^{3/17}
H_y^{-1}~; ~~ E^Y = \xi^Y H_x^{-1}f_x ~y^4 H_y
\end{eqnarray}
Instead, when it is the $D5$-brane the extremal one, $\mu_Y = 0$,
the only solution has a minimally non-extremal $D3$-brane
\begin{eqnarray}
ds^2 &=& H_x^{3/4}H_y^{1/2}[-H_x^{-1}H_y^{-1}f_y dt^2 + H_y^{-1}
(dx^2+ x^2 d\Omega_2^2)+ H_x^{-1}(f_y^{-1} dy^2 + y^2 d\Omega_4^2)
+dw^2] \nonumber\\ e^\phi &=& H_x^{-1/2}~;~~ E^X = \xi^X x^2 H_x
H_y^{-1} f_y~;~~ E^Y = H_x^{-1} y^4 H_y
\end{eqnarray}

\subsubsection*{$(0|D3,NS5|1)$}
It is exactly as in the previous case except for the dilaton,
$e^\phi \to e^{-\phi}$.

\subsubsection*{$(0|D4,D4|1)$}
Let us consider that the first $D4$-brane is the extremal one,
$\mu_X=0$. Then, we have two solutions according to whether the
other one is minimally non-extremal
\begin{eqnarray}
ds^2 &=& H_x^{5/8}H_y^{5/8}[-H_x^{-1}H_y^{-1} f_x dt^2 + H_y^{-1}
(f_x^{-1}dx^2+x^2d\Omega_3^2) + H_x^{-1}(dy^2+y^2 d\Omega_3^2) +
dw^2] \nonumber\\ e^\phi &=& H_x^{-1/4}H_y^{-1/4}~;~~ E^X = x^3
H_x H_y^{-1}~;~~ E^Y = \xi^Y H_x^{-1}f_x ~y^3 H_y
\end{eqnarray}
or a non-minimal non-extremal extension
\begin{eqnarray}
ds^2 &=& H_x^{5/8}H_y^{5/8}[-H_x^{-1}H_y^{-1} f_x dt^2 + H_y^{-1}
f_x^{1/85} (f_x^{-1}dx^2+x^2 d\Omega_3^2) + H_x^{-1} f_x^{1/85}
(dy^2+y^2 d\Omega_3^2) \nonumber\\ && ~~~~~~~~~~~~~ +
f_x^{-6/85}dw^2] \nonumber\\
e^\phi &=& f_x^{-8/85}H_x^{-1/4}H_y^{-1/4} ~;~~ E^X = x^3 H_x
H_y^{-1}~;~~ E^Y = \xi^Y H_x^{-1} f_x ~y^3 H_y
\end{eqnarray}

\bigskip

~

\end{document}